\documentclass[useAMS,usenatbib,usegraphicx,letterpaper]{mn2e}

\usepackage{multirow} 
\usepackage{epsfig}
\usepackage{subfigure}

\newcommand{\etal}{et al.}

\title[The {\it XMM--Newton} view of AGN with IMBH] {The {\it
    XMM--Newton} view of AGN with intermediate mass black holes} \author[G.\ Miniutti \etal]
{G.~Miniutti$^{1,2,3}$\thanks{gminiutti@laeff.inta.es}, G.~Ponti$^2$,
  J.E.~ Greene$^4$, L.C.~Ho$^5$, A.C.~Fabian$^1$, K.~Iwasawa$^{6,7}$ \\ \\
  $^1$Institute of Astronomy, Madingley Road, Cambridge CB3 0HA
  \\
  $^2$ Laboratoire Astroparticule et Cosmologie (APC), UMR~7164, 10 rue A. Domon et L. Duquet, 75205
  Paris Cedex 13, France\\
  $^3$ Laboratorio de Astrof\'isica Espacial y F\'isica Fundamental, LAEFF  (CAB--CSIC--INTA), P.O. Box 78, E--28691, Villanueva de la Ca\~{n}ada, Madrid\\
  $^4$ Princeton University Observatory, Princeton, NY 08544, USA \\
  $^5$ The Observatories of the Carnegie Institution of Washington,
  813 Santa Barbara St., Pasadena, CA 91101, USA\\
  $^6$ Max--Planck--Institut f\"ur extraterrestrische Physik,
  Giessenbachstrasse, 85748 Garching, Germany \\
  $^7$ INAF -- Osservatorio Astronomico di Bologna, Via Ranzani 1, I--40127 Bologna, Italy }

\pagerange{\pageref{firstpage}--\pageref{lastpage}}
\pubyear{2001}

\usepackage{times}

\begin{document}

\label{firstpage}

 \maketitle

\begin{abstract}
  We have observed with {\it XMM--Newton} four radiatively efficient
  active type 1 galaxies with black hole masses $<10^6~M_\odot$,
  selected optically from the Sloan Digital Sky Survey and previously
  detected in the Rosat All Sky Survey. Their X--ray spectra closely
  resemble those of more luminous Seyferts and quasars, powered by
  accretion onto much more massive black holes and none of the objects
  is intrinsically absorbed by cold matter totally covering the
  source. We show here that their soft X--ray spectrum exhibits a soft
  excess with the same characteristics as that observed ubiquitously
  in radio--quiet Seyfert 1 galaxies and type 1 quasars, both in terms
  of temperatures and strength. This is highly surprising because
  the small black hole mass of these objects should lead to a thermal
  disc contribution in the soft X--rays and not in the UV (as for more
  massive objects) with thus a much more prominent soft X--ray
  excess.Moreover, even when the soft X--ray excess is modelled with a
  pure thermal disc, its luminosity turns out to be much lower than
  that expected from accretion theory for the given temperature. 
  While alternative scenarios for the nature of the soft excess
  (namely smeared ionized absorption and disc reflection) cannot be
  distinguished on a pure statistical basis, we point out that the
  absorption model produces a strong correlation between absorbing
  column density and ionization state, which may be difficult to
  interpret and is most likely spurious. Moreover, as pointed out
  before by others, absorption must occur in a fairly relativistic
  wind which is problematic, especially because of the enormous
  implied mass outflow rate. As for reflection, it does only invoke
  standard ingredients of any accretion model for radiatively
  efficient sources such as a hard X--rays source and a relatively
  cold (though partially ionized) accretion disc, and therefore seems
  the natural choice to explain the soft excess in most (if not all)
  cases. The reflection model is also consistent with the additional
  presence of a thermal disc component with the theoretically expected
  temperature (although, again, with smaller--than--expected total
  luminosity).  We also studied in some detail the X--ray variability
  properties of the four objects. The observed active galaxies are
  among the most variable in X--rays and their excess variance is
  among the largest. This is in line with their relatively small black
  hole mass and with expectations from simple power spectra models.
\end{abstract}


\section{Introduction}

Two families of black holes have been observed so far: stellar--mass
black holes are seen in X--ray binaries, while black holes with masses
$10^6$--$10^9~M_\odot$ are ubiquitous in the centers of galaxies,
sometimes revealing themselves as Active Galactic Nuclei (AGN). The
large mass gap between the two families is still scarcely populated.
Dynamical evidence for a black hole with mass $\sim 1.7\times
10^4~M_\odot$ has been reported in the G1 globular cluster (Gebhardt,
Rich \& Ho 2005) and a black hole with a mass of $\sim 4\times
10^4~M_\odot$ may be present in Omega Centauri as
well (Noyola et al. 2008). As for AGN, a limited number of low luminosity
Seyfert--like active galaxies has been shown to harbour
intermediate--mass black holes (IMBH, arbitrarily defined here as
black holes in galaxy centers with $M_{\rm BH} <
10^6~M_\odot$). NGC~4395 is a late--type spiral with no bulge
component and small central stellar velocity dispersion (Filippenko \&
Ho 2003) which exhibits broad permitted optical/UV emission lines, a
compact radio core, and a highly variable point--like X--ray source
unambiguously pointing towards a Seyfert identification (Filippenko,
Ho \& Sargent 1993; Ho \& Ulvestad 2001; Ho et al 2001; Vaughan et al
2005). Mass estimates from the H$\beta$ line width--luminosity
relationship (Filippenko \& Ho 2003) and from X--ray variability
(Vaughan et al 2005) suggest a BH of $10^4-10^5~M_\odot$, consistent
with the upper limit ($10^5~M_\odot$) inferred from the $M$--$\sigma$
relation, but slightly lower than the BH mass of $3.6\times
10^5~M_\odot$, obtained through reverberation mapping (Peterson et
al. 2005). The dwarf elliptical POX~52 also exhibits Seyfert~1 like
broad and narrow emission lines (Kunth, Sargent \& Bothun 1987; Barth
et al. 2004) and the estimate of the BH mass obtained via the H$\beta$
line width--luminosity relationship and the $M$--$\sigma$ relationship
point toward a BH of the order of $10^5~M_\odot$. The main difference
in the AGN properties of the two IMBH is that NGC~4395 is a low
Eddington ratio source ($<10^{-2}$), while POX~52 appears to be
radiating at nearly its maximum rate. Finally, the dwarf Seyfert~1 
SDSS~J160531.84+174826.1 has a relatively broad H$\alpha$ line component
which suggests (via luminosity and line width) a black hole mass of
$\sim 7\times 10^4~M_\odot$ (Dong et al. 2007b).

Greene \& Ho (2004) have defined a sample of 19 IMBH candidates in AGN
using the first data release of the Sloan Digital Sky Survey (SDSS).
Given the adopted selection criteria (low BH mass and prominent
AGN--like spectrum) the sample mostly comprises objects with high
Eddington ratio. The Greene \& Ho work was recently extended to the
fourth data release by Dong et al (2007a) and Greene \& Ho (2007b) who
have increased the sample of IMBH by about one order of
magnitude. Interestingly, about 15 per cent of the larger sample(s)
are suspected to have Eddington ratios below 10 per cent. Due to
selection criteria, and with the exception of NGC~4395, all AGN with
IMBH are actually radiating a very significant fraction of their
Eddington luminosity. The presence of relatively low Eddington ratio
sources in the new sample(s) will make it possible to investigate
NGC~4395--like nuclei in the future, opening a new important window on
IMBH activity.

The original smaller Greene \& Ho sample (hereafter the GH sample) has
been studied in the X--rays by making use of 5~ks long {\it Chandra}
snapshot observations (Greene \& Ho 2007a). The sample is
characterized by X--ray luminosities in the range of
$10^{41}$--$10^{43}$~erg~s$^{-1}$ in the soft 0.5--2~keV X--ray band
and the soft X--ray spectral properties are consistent with those of
more luminous sources harbouring more massive black holes, showing
that black hole mass is certainly not one of the main drivers of soft
X--rays spectral properties. 

Here we report results from {\it XMM--Newton} 40~ks--long observations
of four of the IMBHs from the GH sample which were previously detected
in the Rosat All Sky Survey. The larger effective area and longer
exposure of these observations with respect to the {\it Chandra}
snapshots allow us to consider in some detail the X--ray spectral
properties of these IMBH also above 2~keV, and to compare them with
their more massive and more luminous counterparts (e.g. the PG
quasars).  Moreover, we were able to also study their X--ray
variability properties filling (although with only four objects) a
relatively poorly studied range in black hole masses. Very recently,
the four observations we consider here have been included in a
submitted work by Dewangan et al (2008). Their analysis is mainly
devoted to NGC~4395 and POX~52. As for the common sources, the main
difference with our work is that we focus much more our study on the
spectral analysis and soft excess, exploring different possible models
and interpretations, and comparing the X--ray properties of the four
objects with AGN powered by accretion onto much more massive black
holes.

\section{{\it XMM--Newton} observations}

All observations were performed with the EPIC--pn camera
operated in Full Frame mode. In Table~ 1 we report the source ID (see
Greene \& Ho 2004 for ID definition), redshift, and Galactic column
density. All observations were affected by high background periods.
These were removed before performing spectral analysis and the
resulting net exposure is reported in Table~1. Due to high radiation
affecting the observations of GH~12, the source has been observed
three times (a,b, and c in Table~1 and hereafter). Spectra have been
rebinned so that each energy bin contains at least 20 counts to allow
us to use the $\chi^2$ minimisation technique in spectral fitting.

\begin{table}
\begin{center}
  \caption{Information on the observed sources and observations. The
    short exposure GH~12b does not allow to obtain good quality
    spectra above $\sim$~3~keV and is not considered further.}
\begin{tabular}{lcccc}          
  \hline
  ID &  $\log$M$_{\rm BH}$ & z & $N^{\rm{G}}_H$ & pn exp.\\  
  (1) & (2) & (3) & (4) & (5)\\    
  \hline
  GH~1    & $5.86$&$0.0768$& $3.89$ & 30~ks \\    
  GH~8  &  $5.77$&$0.0811$& $2.34$ & 34~ks \\    
  GH~12a &  $5.98$&$0.106$ & $2.04$ & 17~ks \\    
  GH~12b& -&-&-&  3~ks\\    
  GH~12c& -&-&-& 12~ks\\    
  GH~14 & $5.30$& $0.0281$& $2.04$ & 19~ks\\ 
  \hline   
\end{tabular}
\end{center}
(1) Identification from Greene \& Ho (2004); (2) black hole mass from
Greene \& Ho (2004); (3) redshift; (4) Galactic column density in
units of $\times 10^{20}$~cm$^{-2}$;  (5) Net EPIC--pn exposure used
for the spectral analysis, after filtering for periods of high
background.

\end{table}

\begin{table*}
\begin{center}
  \caption{We apply a broken power law model to all observations in
    the 0.3--10~keV band.  The last column in
    parenthesis gives the fitting result when a single power law model
    is fitted in the whole 0.3--10~keV band. Galactic absorption is
    always assumed. Fluxes (absorbed) are given in units of
    $10^{-13}$~erg~cm$^{-2}$~s$^{-1}$ and luminosities (unabsorbed) in
    units of $10^{43}$~erg~s$^{-1}$. The bolometric luminosity is
    estimated from the optical data as $L_{\rm Bol} = 9 L_{5100}$
    (Greene \& Ho 2004).}
\begin{tabular}{lcccccccc}          
  \hline
  ID & $\Gamma_{\rm s}$ &  $\Gamma_{\rm h}$ &  $F_{0.5-2}$ & $F_{2-10}$&
  $L_{0.5-2}$ &  $L_{2-10}$& $L_{\rm{Bol}}/L_{\rm{Edd}}$ & $\chi^2$/dof \\      
  \hline
  GH~1 & $2.38\pm 0.05$ & $1.8^{+0.4}_{-0.6}$ & $2.58\pm 0.06$ &
  $2.65\pm 0.06$ & $0.43\pm 0.01$ & $0.37\pm 0.01$& $1.12$ &
  308/306 (317/308)\\
%%\hline  
GH~8 & $2.81^{+0.03}_{-0.02}$ & $1.9^{+0.2}_{-0.3}$ & $5.00\pm 0.08$ &
$2.59\pm 0.04$ & $0.93\pm 0.01$ & $0.42\pm 0.01$ & $2.69$ &
401/317 (417/319) \\
%%\hline  
GH~12a & $2.60\pm 0.05$ & $1.9\pm 0.3$ & $3.49\pm 0.08$ & $2.54\pm
0.06$ & $1.12\pm 0.02$ & $0.72\pm 0.02$ & $1.10$ &
178/206 (191/208)\\
%%\hline  
GH~12c & $2.55\pm 0.07$ & $1.7^{+0.4}_{-0.5}$ & $3.4 \pm 0.1$ &
$3.1\pm 0.1$ &
$1.09\pm 0.04$ & $0.85 \pm 0.03$ & $1.10$ &
178/162 (202/164)\\
%%\hline  
GH~14 & $2.0\pm 0.1$ & $1.5\pm 0.2$ & $1.02 \pm 0.04$ & $1.68\pm 0.07$
& $0.019\pm 0.001$ & $0.029\pm 0.001$ & $0.29$ &
74/88 (82/90)\\
\hline
\end{tabular}
\end{center}
%}}
\end{table*}

\section{X--ray spectra}

We first consider the hard 2--10~keV band only, and we apply a simple
power law model (with neutral absorption fixed at the Galactic value)
to all sources and observations. In all cases, we obtain a satisfactory
description of the 2--10~keV spectra although errors on the hard
spectral slope are large (of the order of 20 per cent) due to the poor
signal--to--noise above a few keV in most cases.  The photon index in
the 2--10~keV band is consistent with 1.7--1.9 in all cases, and more
details on the hard spectral slope in the individual objects will be
given below.

\subsection{Fe emission}

The 2--10~keV spectra were also inspected for the presence of iron
(Fe) emission lines. The best evidence for emission lines is seen is
GH~8, where the addition of a narrow ionized Fe emission line (with
energy $6.6 \pm 0.1$~keV and equivalent width $300\pm 200$~eV)
produces a $\Delta\chi^2 = 8$ for 2 degrees of freedom (i.e.
significant at the $\sim$97 per cent level).  An additional narrow
emission line at $7.4\pm 0.2$ (possibly associated with Ni emission)
with an equivalent width of $400\pm 250$~eV is also tentatively
detected (also at the $\sim$97 per cent level, though the large equivalent
width would imply a large Ni abundance, suggesting that the line is
most likely not real).  Due to the low
significance of these emission features, we consider their detection
as tentative only, and we do not discuss them any further (though they are
included in all subsequent fits). In all other objects an additional
narrow emission line in the range 6.4~keV--6.97~keV (neutral to highly
ionized Fe) does not improve the statistics significantly.

However, a narrow ($<50$~eV in $\sigma$) and neutral ($\sim$6.4~keV)
Fe emission line is ubiquitous in the spectra of AGN with typical
equivalent width of $\sim$100~eV (which is also anti--correlated with
X--ray luminosity and Eddington ratio, see e.g. Bianchi et al 2007).
We then include an unresolved Fe K$\alpha$ emission line (with energy
fixed at 6.4~keV) in our spectral models to compute upper limits on
its equivalent width.  We obtain upper limits in the range of
190--550~eV in all objects, which are consistent with expectations
from Bianchi et al. (2007), given the low--luminosity of our objects
(typically a few times $10^{42}$~erg~s$^{-1}$).  Hence, we
do not detect any neutral Fe K$\alpha$ emission line at 6.4~keV, but
the quality of our data is not high enough to rule out that a typical
Fe line is present in the hard X--ray spectra of the sources.
\begin{figure*}
\begin{center}
{\hbox{
\includegraphics[width=0.32\textwidth,height=0.45\textwidth,angle=-90]{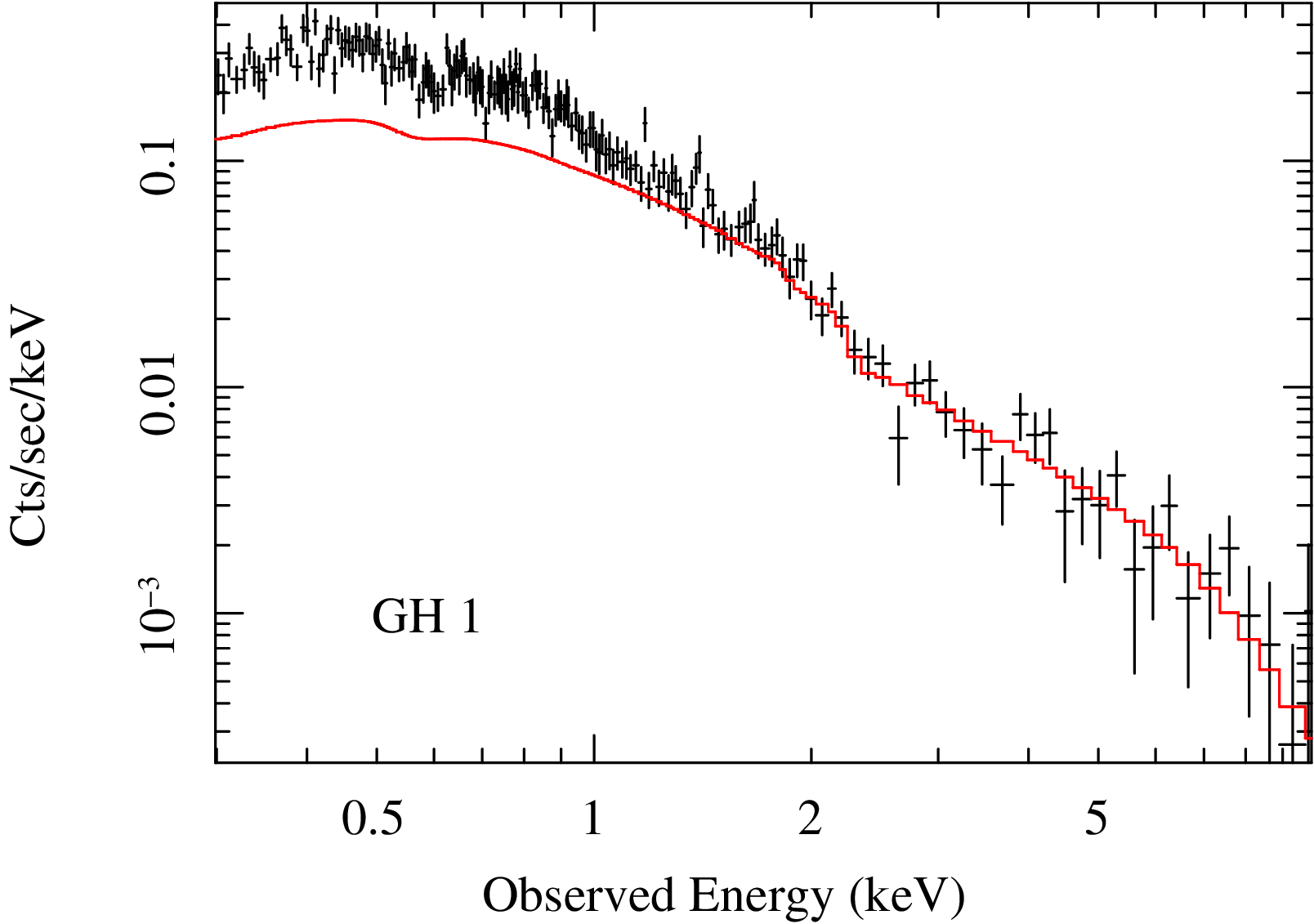}
{\hspace{0.4cm}}
\includegraphics[width=0.32\textwidth,height=0.45\textwidth,angle=-90]{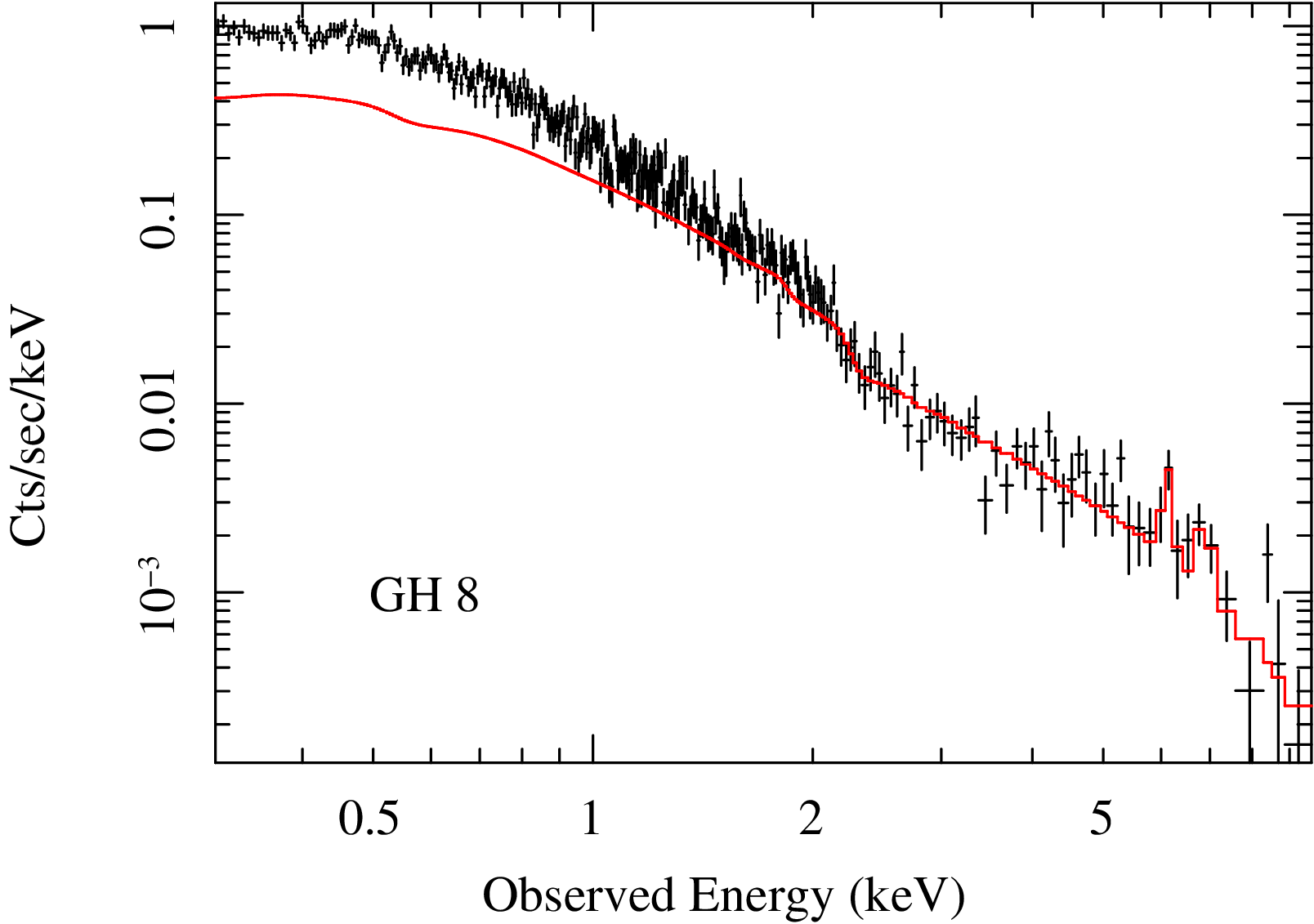}
}}
{\vspace{0.5cm}}
{\hbox{
\includegraphics[width=0.32\textwidth,height=0.45\textwidth,angle=-90]{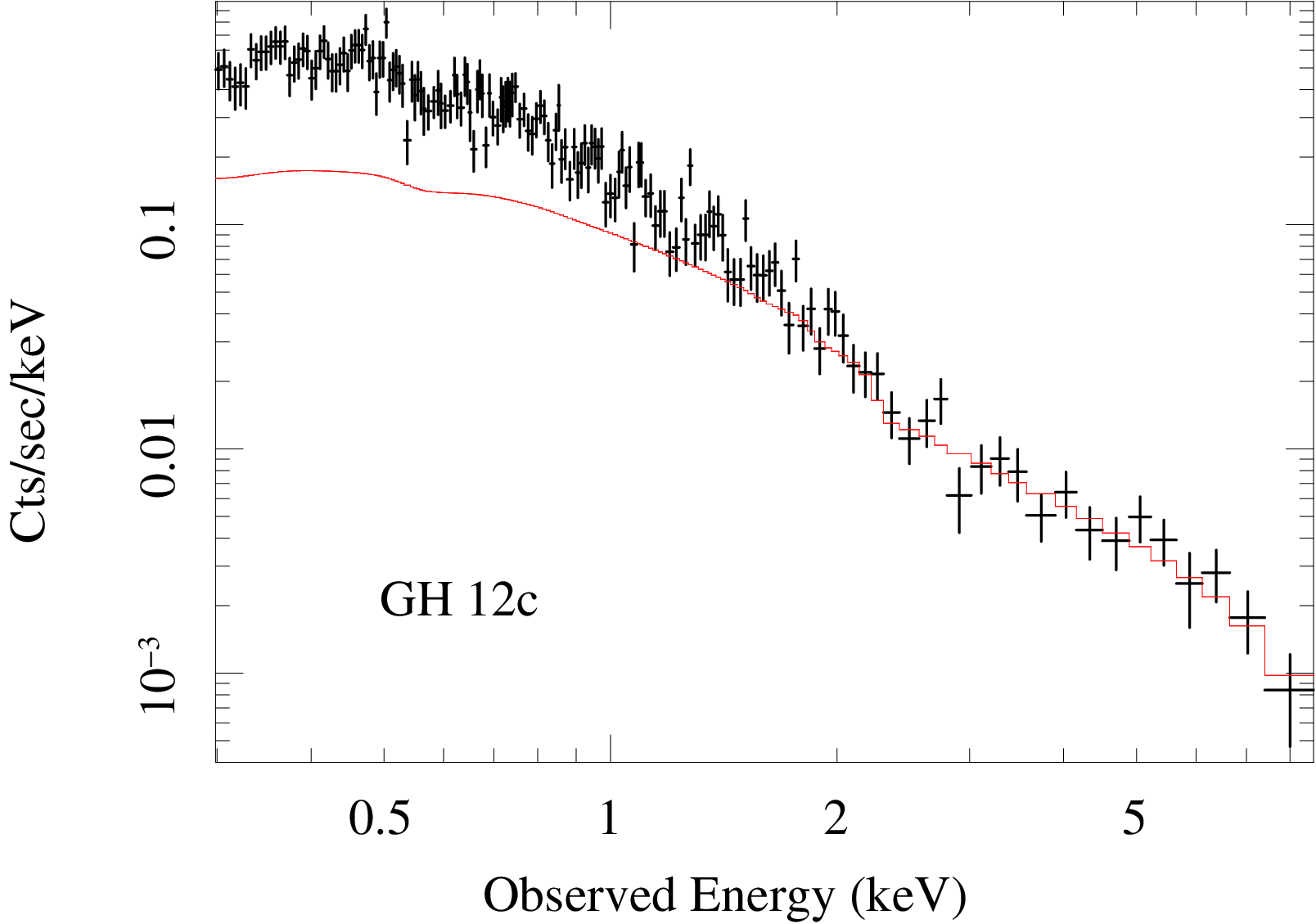}
{\hspace{0.4cm}}
\includegraphics[width=0.32\textwidth,height=0.45\textwidth,angle=-90]{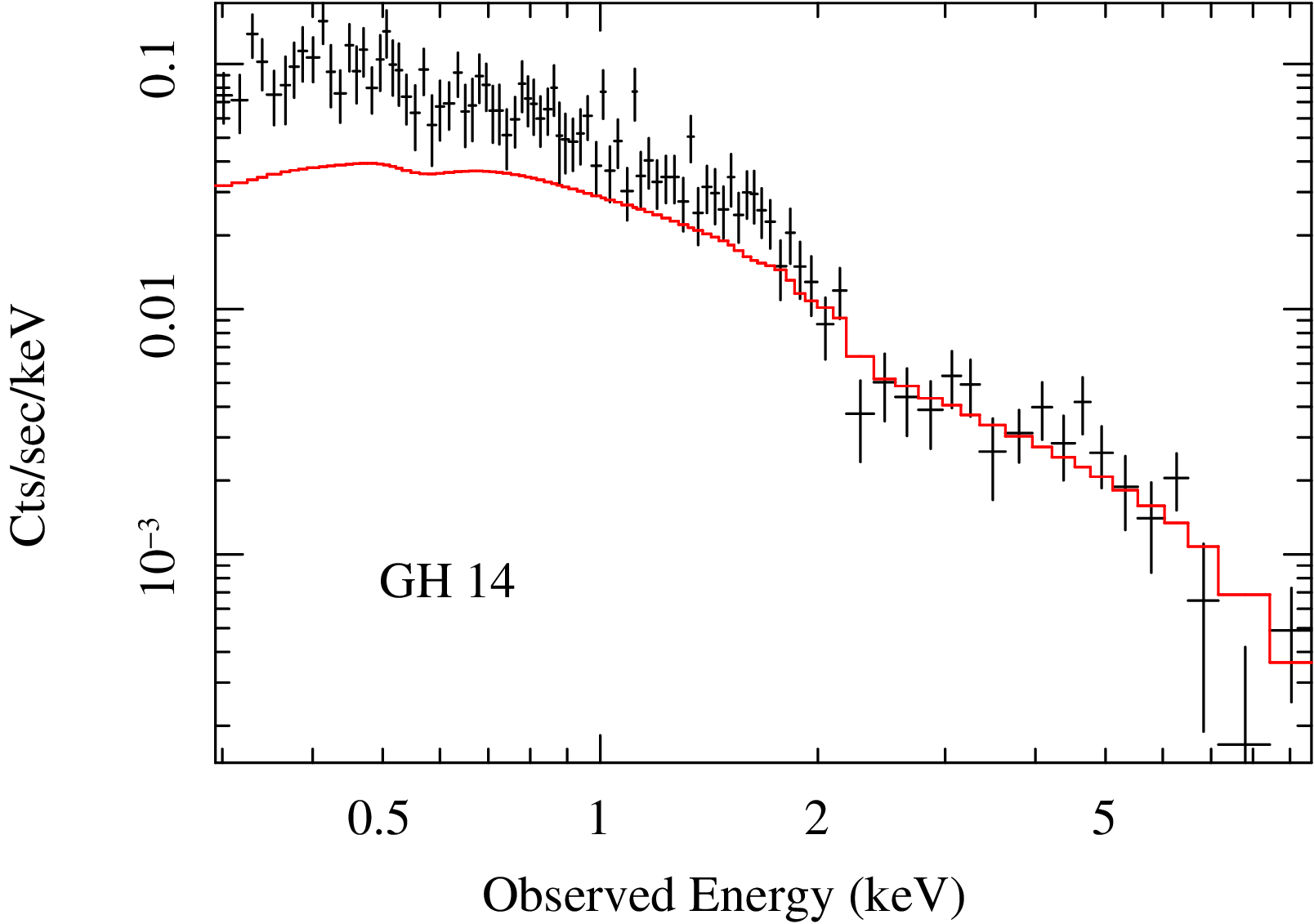}
}}
\caption{The EPIC--pn spectra of the four objects are shown together
  with a power law model fitted in the 2--10~keV band only. In all
  cases, a soft excess is present below 1--2~keV with respect to the
  extrapolated hard spectral model. For source GH~12, we only show one
  observation (GH~12c). When a phenomenological blackbody plus power
  law model is used to fit the spectra (Table~3 and text for details),
  the improvement with respect to the single power law model is at the
  $>$99.99 per cent confidence level in GH~1, GH~8, and GH~12c, 98
  per cent in GH~12a, and is not significant in GH~14.}
\end{center}
\end{figure*}

\subsection{Broadband spectra}

We then extrapolate the hard power law model to soft energies to
assess whether the hard spectral shape provides an adequate
description of the broadband 0.3--10~keV data. In Fig.~1 we show the
result of this exercise which highlights the presence of a soft
excess: the data below 1--2~keV lie in all cases above the
extrapolation of the hard power law model. We then re--fitted the
simple power law model described above in the whole 0.3--10~keV band
and compare it with a (phenomenological) broken power law model. In
Table~2 we report the spectral slopes in the soft ($\Gamma_{\rm s}$)
and hard ($\Gamma_{\rm h}$) band for the broken power law model only
(the break energy is in the 1.5--3~keV range for all sources and is
not reported). The statistics for a single power law model is reported
in parenthesis in the last column and its slope was found consistent
with $\Gamma_{\rm s}$ within the errors due to the much higher
signal--to--noise in the soft band. We also give the derived X--ray
fluxes and luminosities in the 0.5--2~keV and in the 2--10~keV bands
and the Eddington ratio $L_{\rm Bol}/L_{\rm Edd}$ where $L_{\rm Bol} =
9 L_{5100}$ (Greene \& Ho 2004). The spectral fitting results show
that the broken power law model is a better description of the
broadband X--ray spectra of GH~1 (although at the 98.8 per cent level
only), GH~8, and GH~12, while a single power law model is adequate for
GH~14.  In other words, a soft excess is detected in GH~1, GH~8 and
GH~12, and not significantly present in GH~14. We note that the broken
power law model is not entirely adequate to reproduce the soft excess
and that results from such spectral fits should not be used to infer
the significance of the soft excess detection. A blackbody
representation is a better, though also phenomenological,
parametrization of the soft excess (see caption of Fig.~1 and
Table~3). We also point out that in no cases the addition of a neutral
absorption component at the redshift of the source improved the
statistical quality of the fit.  Hence, as already mentioned by Greene
\& Ho (2007a) in their analysis of the {\it Chandra} data, neutral
intrinsic absorption covering the totality of the X--ray source plays
a negligible role in these sources.

\begin{figure}
\begin{center}
\includegraphics[width=0.32\textwidth,height=0.45\textwidth,angle=-90]{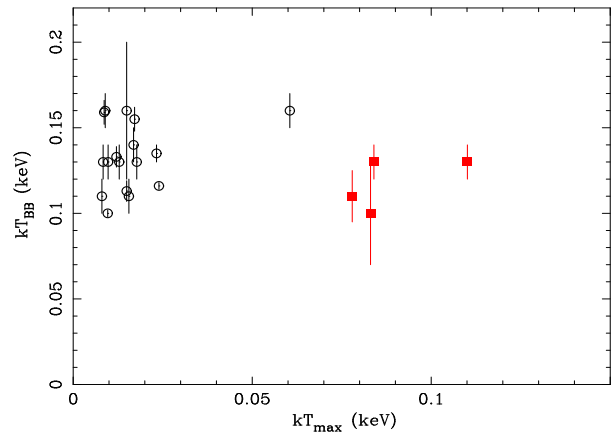}
\caption{The observed temperature of the soft excess is plotted versus
  the maximum temperature expected from the accretion disc
  (Schwarzschild case). Filled squares are for our objects and open
  circles for the sample of radio--quiet PG quasars (results from
  Piconcelli et al 2005). For GH12, the average between observations
  is shown. No correlation is seen, in line with
  previous studies based on the PG quasars.}
\end{center}
\end{figure}

The choice of the broken power law model allows to compare directly
our results on the soft X--ray slope $\Gamma_{\rm s}$ with previous
{\it Chandra} observations of the GH sample (Greene \& Ho 2007a) which
were limited to energies below $\sim$3~keV due to the lack of
signal--to--noise in the hard band (source GH~12 is however observed
only with {\it XMM--Newton}). The soft spectral slopes are in good
agreement with the previous {\it Chandra} observations (Greene \& Ho
2007a), despite long--term flux variability: during the previous {\it
Chandra} observations, GH~1 and GH~14 were 3 and 2 times brighter
(respectively) than in the present {\it XMM--Newton} data, while GH~8
was a factor 1.4 fainter. However, the photon indexes in the
0.5--2~keV band are consistent with each other within the
errors. Moreover, the values of $\Gamma_{\rm s}$ in Table~2 are also
in line with the typical soft X--ray slope of luminous quasars
(e.g. $<\Gamma_{\rm s}> \simeq 2.6-2.7$ in the sample of PG quasars
observed by {\it XMM--Newton} (Porquet et al 2004; Piconcelli et al
2005).  Although affected by large error bars, the hard spectral
slopes are remarkably similar within each other, and also consistent
with the typical 2--10~keV slope of PG quasars ($<\Gamma_{\rm h}>
\simeq 1.9$).  The only exception is represented by the least luminous
source (GH~14), which has a flatter spectral slope in both the soft
and the hard bands with respect to the other objects and, as
mentioned, is also the only source for which a soft excess is not
statistically required. We conclude that the X--ray spectral
properties of our small sample of intermediate--mass black hole AGN do
not significantly differ from the average properties of luminous AGN
powered by accretion onto more massive and more luminous black holes.

\begin{figure}
\begin{center}
\includegraphics[width=0.32\textwidth,height=0.45\textwidth,angle=-90]{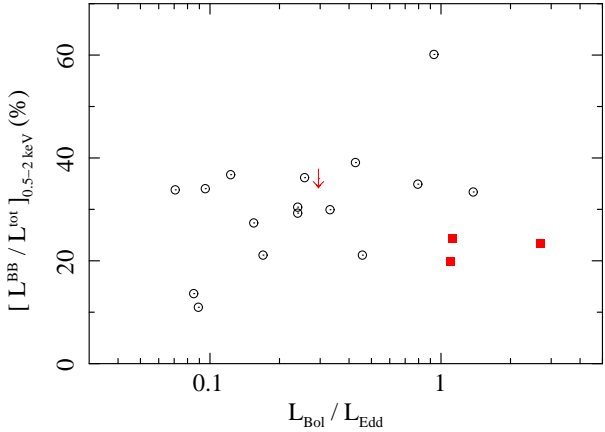}
\caption{The soft excess strength - defined as the ratio between the
  BB and the total luminosity in the 0.5--2~keV band - is plotted
  as a function of Eddington ratio for the radio--quiet PG quasars
  (open circles) and our four IMBH AGNs (filled squares plus the upper
  limit). For GH12, the average between observations is shown. No correlation is seen, and the soft excess strength clusters
  around 30 per cent, though with a large scatter.  }
\end{center}
\end{figure}

\section{The nature of the soft excess}

The nature of the soft excess origin in Seyfert~1 galaxies is still
unclear and matter of debate. As pointed out by several authors, the
idea that the soft excess in AGN represents the high energy tail of
the quasi--blackbody emission from the accretion disc is not
consistent with the observed properties in luminous samples (Czerny et
al 2003; Gierli\'nski \& Done 2004; Crummy et al 2006).  The maximum
temperature for a standard Shakura--Sunyaev thin disc and under the
assuption that the energy release is dominated by thermal disc
emission is achieved at the innermost disc radius $r_{\rm in}$ and
can be rewritten (e.g.  Peterson 1997) as
\begin{equation}
  kT_{\rm{max}} \simeq 23.8~~M_8^{-1/4}~(\dot{M}/\dot M_{\rm
    Edd})^{1/4}~(r_{\rm in}/6~r_g)^{-3/4}\rm{eV}
  \label{kt}
\end{equation}
where $M_8$ is the black hole mass in units of $10^8~M_\odot$ and
$r_g=GM/c^2$. The minimum possible inner disc radius $r_{\rm in}$ is
generally assumed to coincide with the last stable circular orbit of a
test particle orbiting the black hole in the equatorial plane and
ranges from 1.24~$r_g$ to 6~$r_g$ (for a maximally rotating Kerr black
hole and a non--rotating Schwarzschild one respectively). 

For the typical black hole masses and accretion rates of PG quasars
and a standard optically thick and geometrically thin accretion disc,
one would expect an average maximum disc temperature ($kT_{\rm max}$)
of the order of 20~eV (assuming a Scwarzschild hole), with a
dispersion of about one order of magnitude (and even smaller
temperatures if the energy release is not completely dominated by
thermal emission).  However, the observed temperature of the soft
excess in PG quasars clusters with very little dispersion around
100--150~eV, too hot and too uniform to be associated with disc
emission (Porquet et al 2004; Piconcelli et al 2005; Brocksopp et al
2006). Also, when highly variable sources are examined, the soft
excess luminosity does not scale with temperature as expected naively
from the Planck law (see e.g. Ponti et al 2006).  For the IMBH objects
we are considering here, the expected $kT_{\rm max}$ ranges between
80~eV and 110~eV for a Scwarzschild hole (and a factor $\sim 3.26$
higher for a Kerr black hole), and the blackbody emission should
largely enters the soft X--ray regime accessible with our {\it
XMM--Newton} observations, so that the soft X--ray spectra of AGN with
IMBH and of (say) PG quasars should significantly differ.
\begin{figure}
\begin{center}
\includegraphics[width=0.32\textwidth,height=0.45\textwidth,angle=-90]{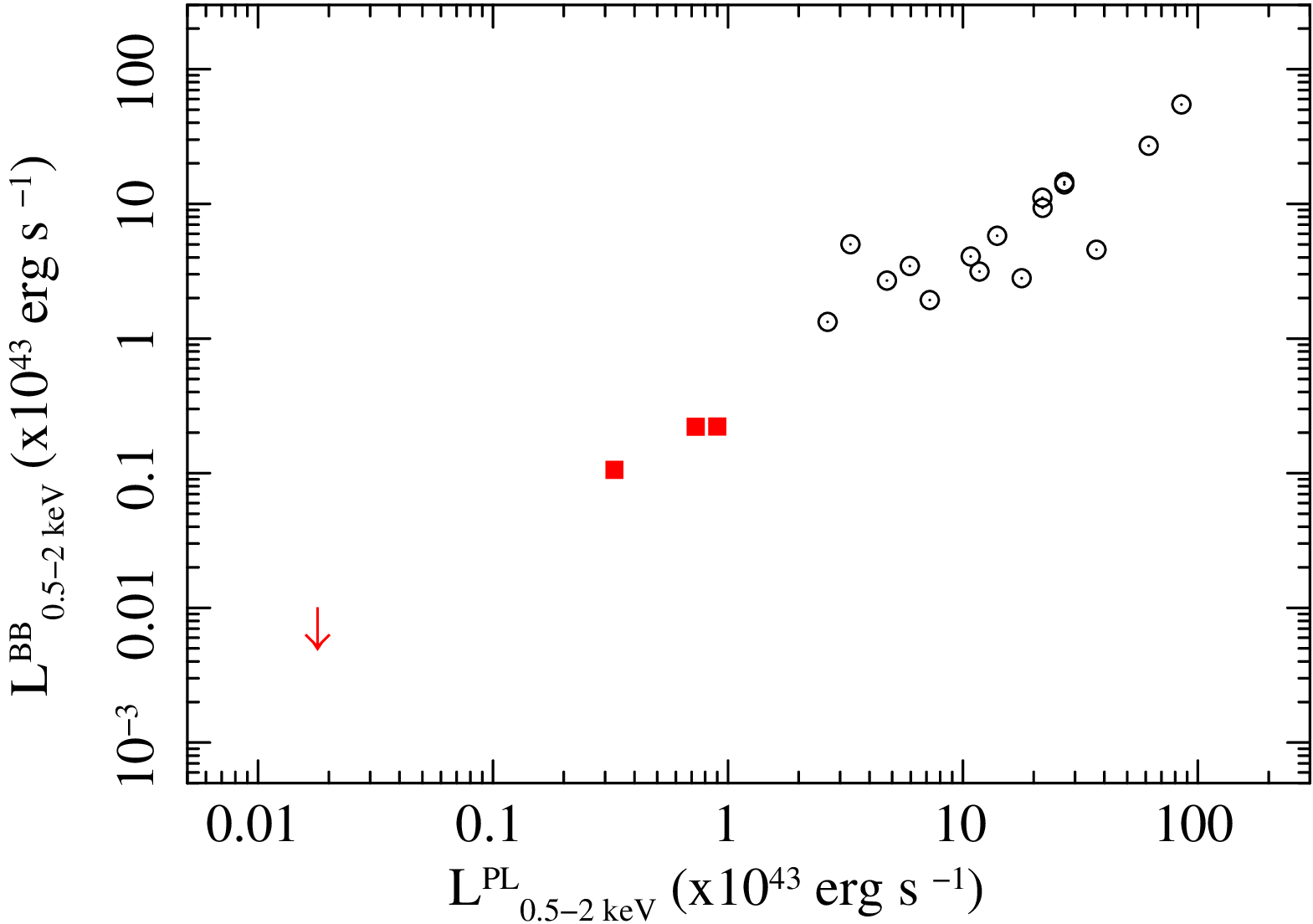}
{\vspace{0.5cm}}
\includegraphics[width=0.32\textwidth,height=0.45\textwidth,angle=-90]{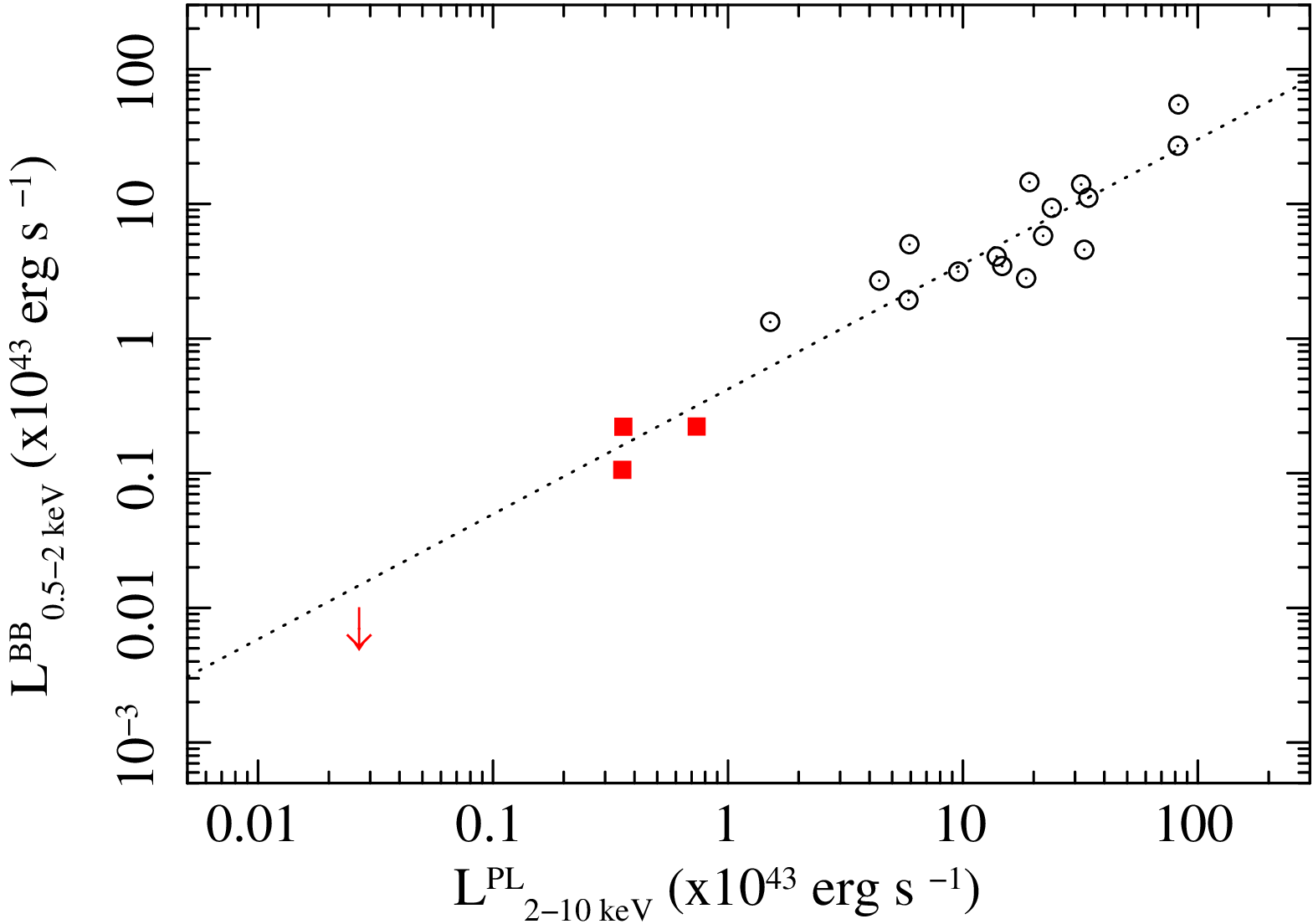}
\caption{Results from phenomenological BB plus PL fits of the X--ray
  spectrum for the radio--quiet PG quasars and for our IMBHs (filled
  squares and upper limit). For GH~12, we plot the average between
  observations 12a and 12c. In the top panel we show the correlation
  between the 0.5--2~keV BB luminosity and the PL luminosity in the
  same band. The BB luminosity is also correlated with the PL
  luminosity in the hard 2--10~keV band, as shown in the bottom panel
  ($\log L^{\rm{BB}}_{0.5-2~keV} = 0.99\times \log
  L^{\rm{PL}}_{2-10~keV}$).}
\end{center}
\end{figure}

However, as mentioned above, the soft photon index $\Gamma_{\rm s}$ we
measure is consistent with the average soft photon index in PG
quasars, pointing towards a similar spectral shape. This is confirmed
when the soft excess in our objects is modeled with a blackbody
component. In this case, the inferred temperature turns out to be
$\sim$~100--150~eV (see Table~3), remarkably similar to that measured
from identical fits to PG quasars (Piconcelli et al 2005). This is
shown in Fig.~2 where we plot the observed temperature as a function
of the theoretical maximum temperature expected from an accretion disc
around a Schwarzschild black hole. Although 100--150~eV is a
reasonable value for the disc temperature around our
intermediate--mass objects, it is quite remarkable that the inferred
temperature is that similar to that of PG quasars, casting doubts on
its interpretation in terms of thermal emission, and pointing towards
a common and different origin for the soft excess in radiatively
efficient black holes of all masses. Moreover, even if all soft
X--ray emission is attributed to a thermal disc spectrum, the inferred
blackbody luminosity is, for all sources, much lower than the expected
one (which can be obtained e.g. by applying the Planck low to
Eq.~\ref{kt}). Taking as a representative example GH~12, a
blackbody--only fit in the 0.3--0.9~keV band reasonably describes the
data ($\chi^2 /dof = 250/197$) with a temperature of $\sim 137$~eV for
a total blackbody luminosity of $\sim 1.7\times 10^{43}$~erg~s$^{-1}$,
well below that expected one for the given black hole mass and mass
accretion rate (in excess of $10^{44}$~erg~s$^{-1}$).

\begin{table}
\begin{center}
  \caption{Results of spectral fits in which the broadband X--ray
    spectra are 
    modeled with a blackbody (BB) plus power law model (with photon
    index $\Gamma_h$). For GH~14, the soft excess is not
    statistically required.}
\begin{tabular}{lcccc} 
  \hline
  \multicolumn{2}{l}{BB} & \\
  \hline
  ID & $T_{\rm BB}$ &  SE$^\dagger$ (\%) & $\Gamma_h$ & $\chi^2$/dof \\     
  \hline 
  GH~1 & $0.13\pm 0.01$ & $24.4\pm 6.0$&$2.05\pm 0.08$ & 293/306\\
  GH~8 & $0.13\pm 0.01$ & $23.3\pm 4.0$&$2.5\pm 0.1$ & 346/317\\
  GH~12a & $0.11\pm 0.02$ &$16.8\pm 6.0$& $2.4\pm 0.1$ & 184/206\\
  GH~12c & $0.11\pm 0.01$ &$23.0\pm 6.0$& $2.1\pm 0.1$ & 175/162\\
  GH~14 & $0.10\pm 0.03$ &$<36$& $1.8\pm 0.2$ & 77/88\\
  \hline
\end{tabular}
\parbox{3.28in} {
$^\dagger$ soft excess (SE) strength in the 0.5--2~keV band:
this is the ratio of the BB luminosity to the total one
in the 0.5--2~keV band, expressed as a percentage.
} 
\end{center}
\end{table}

As mentioned above, the only source with no need for a soft excess
(GH~14) is also the least luminous, in both absolute and Eddington
ratio terms ($L_{\rm Bol}/L_{\rm Edd} \sim 0.3$, see Table~2). When
fitting the X--ray spectra with a blackbody plus power law model, the
strength of the soft excess can be (phenomenologically) defined as the
contribution of the blackbody component to the total luminosity in the
soft (0.5--2~keV) band. In the case of GH~14, the blackbody
contribution is only an upper limit of 36 per cent (Table~3). In order
to see if the soft excess is more prominent in higher Eddington ratio
sources, we computed the contribution of the blackbody component to
the 0.5--2~keV luminosity in a sample of PG quasars (fitting results
from Piconcelli et al 2005) and in our intermediate--mass black hole
AGN, and we plot it as a function of Eddington ratio in Fig.~2. As can
be seen, no correlation with Eddington ratio is seen, and the
blackbody contribution in the 0.5--2~keV band clusters around 30 per
cent (although with significant scatter), despite more than one order
of magnitude variation in Eddington ratio, i.e. the soft excess
strength is nearly uniform in the sample (see e.g. Fig.~4 in
Middleton, Done \& Gierli\'nski 2007). 

The apparent uniform $\sim$~30 per cent contribution of the soft
excess to the soft X--ray luminosity means that the BB and PL
luminosities are reasonably well correlated in objects with different
black hole mass and mass accretion rate. This is indeed shown in
Fig.~3 (top) where we plot the BB luminosity versus the PL one in the
0.5--2~keV band (as inferred from BB plus PL fits to the X--ray
spectra of PG quasars, from Piconcelli et al. 2005, and of our IMBH).
Although the conversion between the PL 0.5--2~keV and 2--10~keV
luminosities depends on the PL slope and is hence object--dependent,
it is obvious, that $L^{\rm BB}_{0.5-2~keV}$ is also correlated with
the 2--10~keV PL luminosity (see bottom panel of Fig.~3). The good
correlation means that the soft excess strength can be used to predict
with good accuracy the 2--10~keV luminosity (and vice--versa) implying a
physical link between the hard power and the soft excess. Our IMBH are
no different from their more massive counterparts and they extend the
relationship to lower luminosities (and black hole masses).

\subsection{Alternative models I: the smeared absorption model}

The universal shape of the soft excess could be better explained if it
originated, at least partially,  from processes that are genuinely independent from the
black hole mass and the AGN luminosity, or accretion rate. Atomic
processes can give rise to absorption and emission in the X--ray band
affecting the overall spectral shape, and thus provide a possible
alternative to optically thick thermal emission as an explanation for
the X--ray soft excess. 

For instance, Gierli\'nski \& Done (2004)
proposed a model in which the soft excess is an artifact due to
absorption of an underlying simple power law continuum. The model
requires a moderately steep intrinsic X--ray continuum
($\Gamma=2\div3$) and absorption by a substantial column of ionized
gas, which mostly affects the 0.7--3~keV band, producing an apparent
soft excess. Since sharp absorption features are absent in the X-ray
spectra, the model also does require very significant smearing which
may be due to large velocities of the absorbing gas, as could be
found, for instance, in a wind originating from the innermost
accretion flow. The model potentially reproduces the X-ray variability
properties of some sources (see Gierli\'nski \& Done 2006), although
flat rms spectra in highly variable sources seem difficult to account
for (e.g. during one {\it XMM--Newton} observations of 1H~0707--495,
see Fig.~4 in Boller et al 2003; and also in Mrk~335 see O'Neill et al
2007 and Larsson et al 2008).

\begin{figure}
\begin{center}
\includegraphics[width=0.32\textwidth,height=0.45\textwidth,angle=-90]{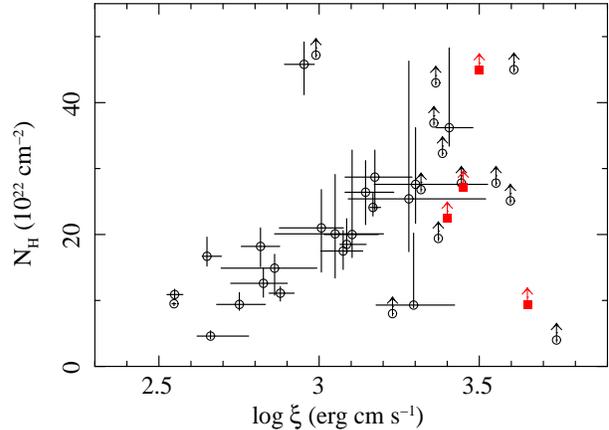}
\caption{The correlation between column density and ionization
  parameter as inferred from fitting the smeared absorption model to a
  sample of PG quasars and NLS1s (open circles results from Middleton,
  Done \& Gierli\'nski 2007) and to our four IMBHs (filled squares,
  for GH~12 only the first observation is shown for simplicity, see
  Table~4).  Results for which the column density upper limit is equal
  or larger than the highest tabulated value of $5\times
  10^{23}$~cm$^{-2}$ are treated as lower limits at the best--fitting
  ionization parameter.  }
\end{center}
\end{figure}

When applied to samples of PG quasars and NLS1s, the model provides
reasonable spectral fits to the X--ray spectra and requires column
densities $N_H\sim1\div 5\times 10^{23}$~cm$^{-2}$ with ionization
$\log\xi\sim2.5\div 3.5$ (Middleton, Done \& Gierli\'nski 2007).  For
the same ionization, larger columns lead to apparent flatter hard
spectra so that two sources with identical intrinsic spectrum but
different column densities produce very different soft excess
strength (in contrast with Fig.~2 and also with Fig.~4 in Middleton,
Done \& Gierli\'nski 2007).

The only way out for the model to produce a uniform spectral shape and
soft excess strength is to force the ionization parameter to be higher
for larger column densities, so to reduce the absorption trough and
produce a more uniform soft excess strength for different column
densities.  It is then not surprising that, when the smeared
absorption model is fitted to the X--ray spectra of a sample of PG
quasars and NLS1 galaxies, a tight correlation between ionization
parameter and column density is observed. In Fig.~5 we show the
$N_H-\xi$ correlation as inferred from fitting the model to PG quasars
and NLS1s (from Middleton, Done \& Gierli\'nski 2007) and to our four
IMBHs (results for our objects are reported in Table~4). As mentioned,
the correlation of Fig.~5 is most likely produced  because the model tries
to reproduce the uniform spectral shape and soft excess strength by
adjusting the two parameters.

Hence, a question arises: is the $N_H-\xi$ correlation physical in
origin, or is it just driven by the model attempting to reproduce the
spectral shape by adjusting the parameters in a non physical manner?
Simple physical arguments imply that the correlation is not naturally
produced: the ionization parameter is defined as $\xi = L/(nr^2)$
where $L$ is the nuclear luminosity, $n$ the absorber density, and $r$
its distance from the nuclear photo--ionizing source.  If the absorber
is a shell of thickness $\Delta r$, then $\xi = L/(rN_H)~\Delta r/r$.
Since both luminosity and size are expected to roughly scale with the
black hole mass, $\xi$ and $N_H$ are naively expected to be
anti--correlated (if correlated at all). It is also possible that the
correlation seen in Fig~5 is simply due to a degeneracy in the
parameter space of the model. If so, and given that the only other
parameter (the velocity smearing) is also largely unconstrained (and
generally only a lower limit), the model does not provide significant
insights on the properties of the absorber (i.e. any column density
will be able to provide a good parameterization provided that the
ionization state lies on the $N_H-\xi$ correlation of Fig.~5).

\subsection{Alternative models II: the disc reflection model}

On the other hand, the soft excess universal shape may be due to the
presence of an atomic--related emission component rather than
absorption. The natural candidate is X--ray reflection from the
irradiated accretion disc in which the sharp atomic features (mainly
fluorescent emission lines and absorption edges from the most abundant
metals) are broadened and smeared out quite naturally because of the
high velocities and strong gravity effects in the inner disc. The
model has the advantage that such a spectrum has been observed with
very little ambiguity in some AGN (see Fabian \& Miniutti 2007 for a
review) and Galactic X--ray binaries (see Miller 2007 for a review).
As the absorption model, disc reflection potentially
explains not only the spectral shape, but also the spectral
variability properties of many sources (Miniutti \& Fabian 2004;
Fabian et al 2004; 2005; Ponti et al 2006; Miniutti et al. 2007). The
{\it XMM--Newton} bandpass is too limited to disentangle statistically
between reflection and absorption, which predict different spectral
shapes mainly above 10~keV. We provide a comparison in Table~4, showing that
the two models are indeed indistinguishable in terms of fitting
statistics.
\begin{table}
\begin{center}
  \caption{Results of spectral fits in the 0.3--10~keV band with the
    two alternative models for the soft excess. {\it Smeared
    Absorption:} column densities are given in units of
    $10^{22}$~cm$^{-2}$. The velocity dispersion $\sigma$ is a free
    parameter of the model but is not reported here because it is
    always a lower limit of at least 0.25~c. {\it Disc Reflection:}
    The disc inclination is allowed to vary in the range of $0^\circ -
    45^\circ$ only (to be consistent with the Seyfert~1 nature of the
    sources). The disc extends from the last stable orbit for
    Schwarzschild or Kerr black holes out to $400~r_g$ and the
    best--fitting case is reported in the last column (K or S). The
    emissivity of the reflecting disc is fixed to the standard
    $r^{-3}$ profile and solar abundances are assumed. In the first
    column, we report the reflection fraction of the disc reflection
    component. In the bottom part of the table, we present a
    model in which we add thermal disc emission to the reflection
    model ({\it Disc Reflection + Disc BB}). Though not required
    statistically with high significance, this solution is more
    self--consistent than that with reflection only (see text). We use
    the {\small DISKBB} model, and its temperature is set to the
    maximum one for a Schwarzschild or a Kerr black hole (according to
    Eq.~\ref{kt}). The geometry is given in parenthesis and represents
    the best--fitting case from the simultaneous constraints of the
    thermal and reflection models (we adopt Schwarzschild for
    GH~14). Part of the soft excess is now described by thermal
    emission, and we report the ratio of the BB--to--reflection flux
    in the 0.5--2~keV band for comparison.}
\begin{tabular}{lcccc} 
  \hline
  \multicolumn{2}{l}{Smeared Absorption} & \\
  \hline
  ID & $N^{(a)}_H$ &  $\log \xi$ & $\Gamma$ & $\chi^2$/dof \\    
  \hline 
  GH~1 & $\geq 27$ & $3.45\pm 0.25$&$2.11\pm 0.08$ & 301/305\\
  GH~8 & $\geq 45$ & $3.50\pm 0.20$&$2.59\pm 0.07$ &376/316\\
  GH~12a & $\geq 22$ &$3.40 \pm 0.25$& $2.45\pm 0.15$ &178/205\\
  GH~12c & $\geq 35$ &$3.52 \pm 0.28$& $2.30\pm 0.19$ &172/161\\
  GH~14 & $\geq 10 $ &$\geq 3.0$& $1.80\pm 0.20$ &76/88\\
  \hline
  \multicolumn{2}{l}{Disc Reflection} & \\
  \hline
  ID & $R$ &  $\log \xi^{(b)}$ & $\Gamma$ &$\chi^2$/dof \\     
  \hline 
  GH~1 & $1.2\pm 0.5$ & $1.98\pm 0.25$ &$2.20\pm 0.09$ & 296/305 (K)\\
  GH~8 & $1.9\pm 0.4$ & $2.48\pm 0.30$ &$2.60\pm 0.14$ & 350/316 (S)\\
  GH~12a & $0.9\pm 0.6$ &$2.46 \pm 0.35$& $2.42\pm 0.16$ &179/205
  (K)\\
  GH~12c & $1.2\pm 0.6$ &$2.66 \pm 0.35$& $2.20\pm 0.17$ &177/161
  (K)\\
 GH~14 & $\leq 0.8 $ & $-$ & $1.82\pm 0.20$ &78/88 (K/S)\\
  \hline
  \multicolumn{2}{l}{Disc Reflection + Disc BB} & \\
  \hline
  ID & $kT^f$ &  BB/Ref$^{(c)}$ & $\Gamma$ &$\chi^2$/dof \\     
  \hline 
  GH~1 & $0.275$ & $1.63 \pm 0.65$ &$2.26\pm 0.11$ & 293/304 (K)\\
  GH~8 & $0.356$ & $0.23\pm 0.16$ &$2.70\pm 0.18$ & 346/315 (K)\\
  GH~12a & $0.255$ &$< 1.90$& $2.48\pm 0.19$ &178/204
  (K)\\
  GH~12c & $0.255$ &$< 1.60$& $2.30\pm 0.25$ &173/160
  (K)\\
 GH~14 & $0.083$ & $-$ & $1.82\pm 0.20$ &77/87 (S)\\
\hline
\end{tabular}
\parbox{3.28in} { $^{(a)}$ The highest tabulated value in the model is
$N_H = 50\times 10^{22}$~cm$^{-2}$. Lower limits mean that the
parameter could not be constrained below that value. $^{(b)}$ The
ionization state of the reflector is unconstrained in GH~14. $^{(c)}$
The ratio between the BB and the reflection flux in the 0.5--2~keV
band is given to compare the contribution of these two
components. The ratio and spacetime geometry are unconstrained in
GH~14, given that the soft excess itself in this object is an upper
limit only.}
\end{center}
\end{table}

However, the two models differ in their physical interpretation and
consequences.  The disc reflection model main parameters are i) the
ionization state of the reflector and ii) the reflection fraction,
i.e.  the relative strength of the reflection component with respect
to the intrinsic continuum. The inclination of the line of sight with
respect to the disc also plays an important role affecting the overall
spectral shape (especially if it is larger than 50 degrees), but we
assume in the qualitative discussion below that the inclination is
smaller for all Seyfert 1 galaxies.

To be successful, the model needs to explain the uniform
spectral shape and soft excess strength in different sources by
adjusting these parameters in a physically reasonable way.  Generally
speaking, a larger reflection fraction (for fixed $\xi$) produces
flatter hard spectral shapes (with a similar effect as the increase in
column density in the smeared absorption model). Hence, if the
ionization parameter stays the same, the model runs into the same
problems as the smeared absorption one, and is unlikely to be able to
produce a uniform soft excess strength. 

However, a correlation between ionization state of the disc and
reflection fraction is qualitatively naturally expected in the reflection
interpretation. Indeed, a larger reflection fraction means that the
continuum irradiates preferentially the disc rather than the observer
or, in other words, that the disc subtends a larger solid angle at the
X--ray continuum source.  The inner disc irradiation is therefore
stronger for higher reflection fractions and one expects, for a given
disc density, an increase of the ionization of the inner disc.  

If we assume that the correlation is linear, i.e. that a doubling of
the reflection fraction induces a doubling of the ionization
parameter, we can test whether this correlation would give rise to a
uniform soft excess. We simulate two 100~ks {\it
  XMM--Newton} spectra which differ in reflection fraction ($R=1$ and
$R=2$ respectively) and disc ionization ($\xi = 250$~erg~cm~s$^{-1}$
and $\xi =500$~erg~cm~s$^{-1}$ respectively). By fitting the two
spectra with a blackbody plus power law model, we obtain a soft excess
strength of $\sim$22 and $\sim$30 per cent respectively, in line with
the spread and absolute value observed in Fig.~2. In other words,
doubling the reflection fraction and the ionization parameter only
produces a marginal increase in the soft excess strength, fully
consistent with the observed spread.

\begin{figure}
\begin{center}
\includegraphics[width=0.32\textwidth,height=0.45\textwidth,angle=-90]{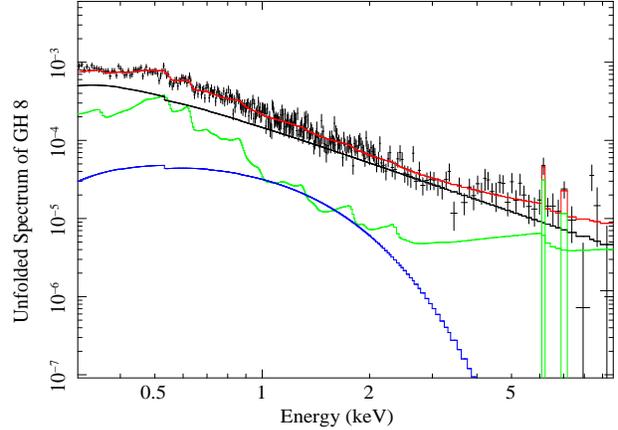}
\caption{The unfolded spectrum and model components for the Reflection + Disc BB fit of GH~8.}
\end{center}
\end{figure}

\subsubsection{Adding the disc thermal emission component}

In contrast with the case of large black hole mass AGN where thermal
  disc emission is in the UV, for the black hole masses and accretion
  rates (Tables~1 and 2) of our mini--sample of IMBHs, it should
  significantly contribute at soft X--ray energies especially if the
  disc extends down to its last stable orbit around the black hole
  (see Eq.~\ref{kt}). Then, even if reflection alone can account for
  all the soft excess, a self--consistent model should also include
  thermal disc emission. We then added a disc blackbody component (the
  multicolor {\small DISKBB} model) with temperature fixed to that
  expected from the standard thin disc model (Eq.~\ref{kt}) in the two
  extreme cases of Schwarzschild and maximally spinning Kerr
  geometries. The relativistic blurring of the reflection component is
  self--consistently computed according to the two spacetime
  geometries and we report results for the best--fitting geometry
  only. In one case (GH~8), the addition of the thermal component
  suggests a change (from Schwarzschild to Kerr) of the best--fitting
  geometry. The results are presented in the bottom part of Table~4
  and show that, though not required formally by the data, a thermal
  disc component with the right temperature is consistent with the
  X--ray spectra of our IMBHs, even when the reflection component is
  introduced. 

We point out however that the inferred blackbody luminosity is
  again (see Section~4) far below that expected from accretion
  theory. This is not completely surprising given that the estimate in
  Eq.~\ref{kt} assumes, by definition, that all of the accretion
  energy release is in the form of a thermal component, i.e. that the
  expected blackbody luminosity is comporable to the bolometric. This
  may be accurate for BH binaries in disc--dominated (or soft) states,
  but in the present case of small black hole mass AGN, and even if
  the thermal component should dominate the soft X--rays, much of the
  X--ray emission is clearly not in the form of a thermal component,
  which makes it difficult to predict reliably the expected blackbody
  temperatures and luminosities. It should also be stressed that the
  low inferred blackbody luminosity is not an artifact of the presence
  of the reflection component eating away part of the soft
  excess. Even if the soft excess is modelled entirely as a blackbody,
  its total unabsorbed luminosity is in all of our sources more than
  one order of magnitude below the expected one (see Section~4). To
  illustrate the relative contribution of the thermal (blue) and
  reflection (green) components, we show the unfolded spectrum and
  model components for the case of GH~8 in Fig.~6.

\begin{figure*}
\begin{center}
{\hbox{
\includegraphics[width=0.35\textwidth,height=0.47\textwidth,angle=-90]{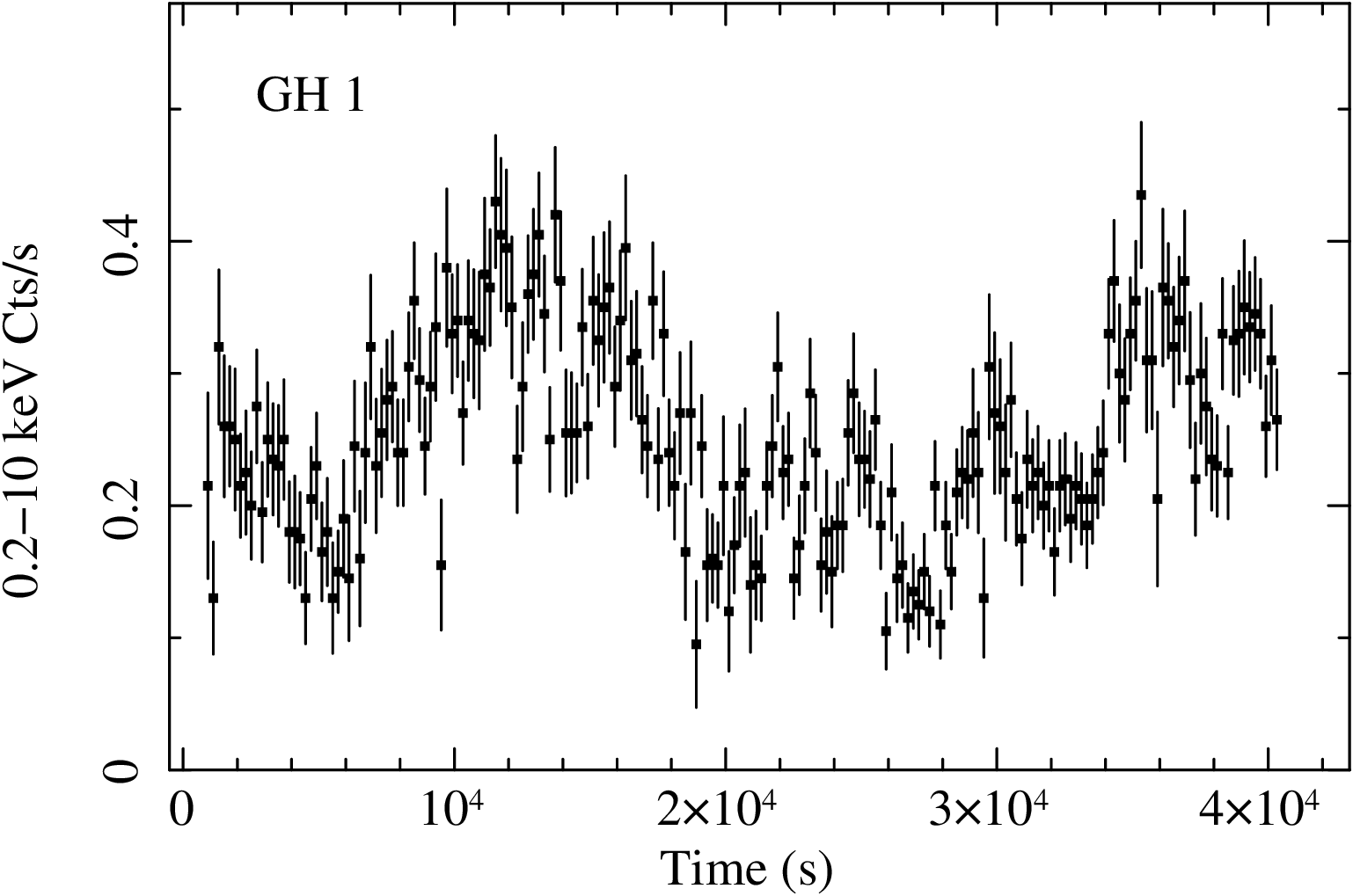}
{\hspace{0.4cm}}
\includegraphics[width=0.35\textwidth,height=0.47\textwidth,angle=-90]{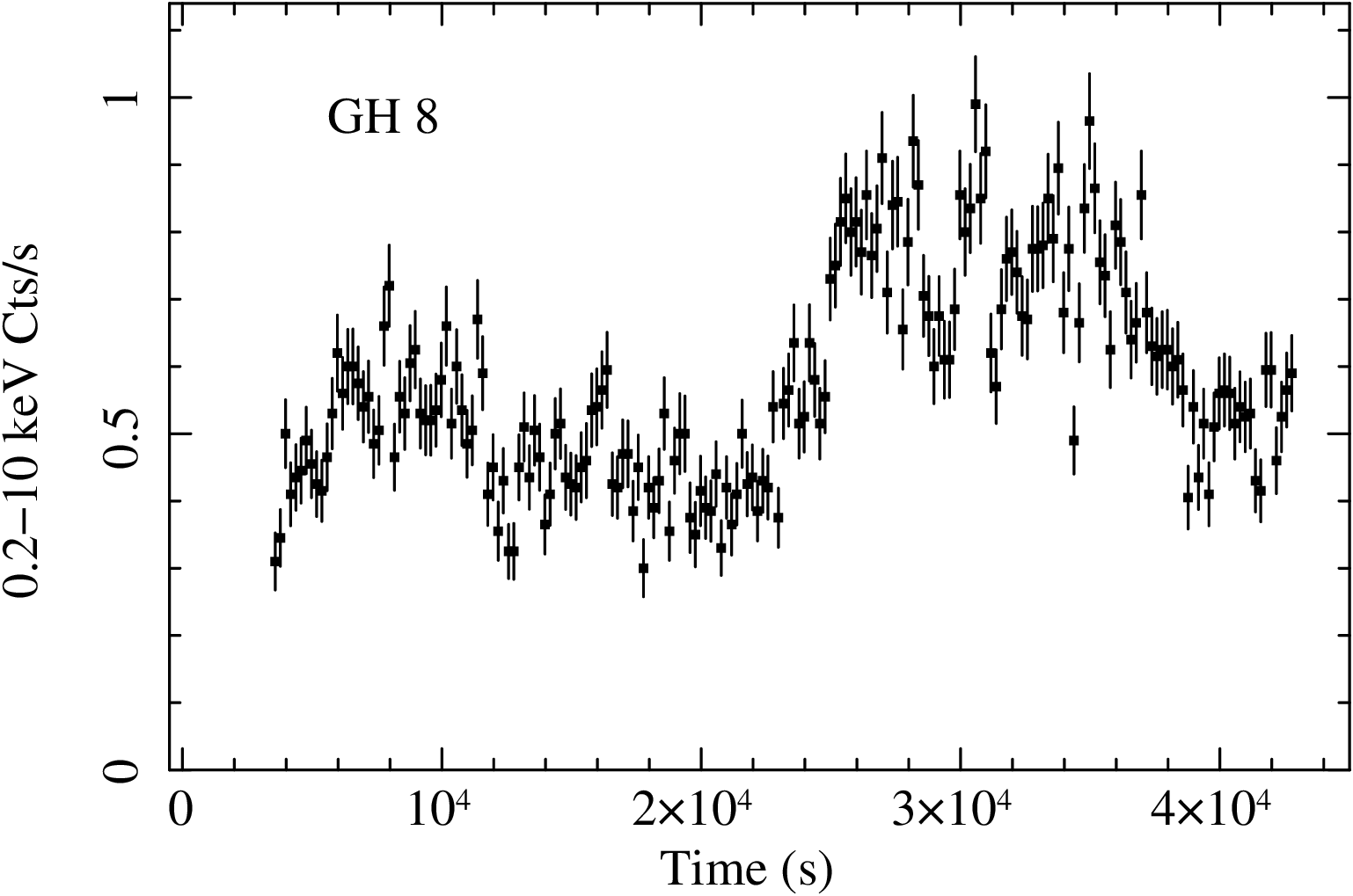}
}}
{\vspace{0.5cm}}
{\hbox{
\includegraphics[width=0.35\textwidth,height=0.305\textwidth,angle=-90]{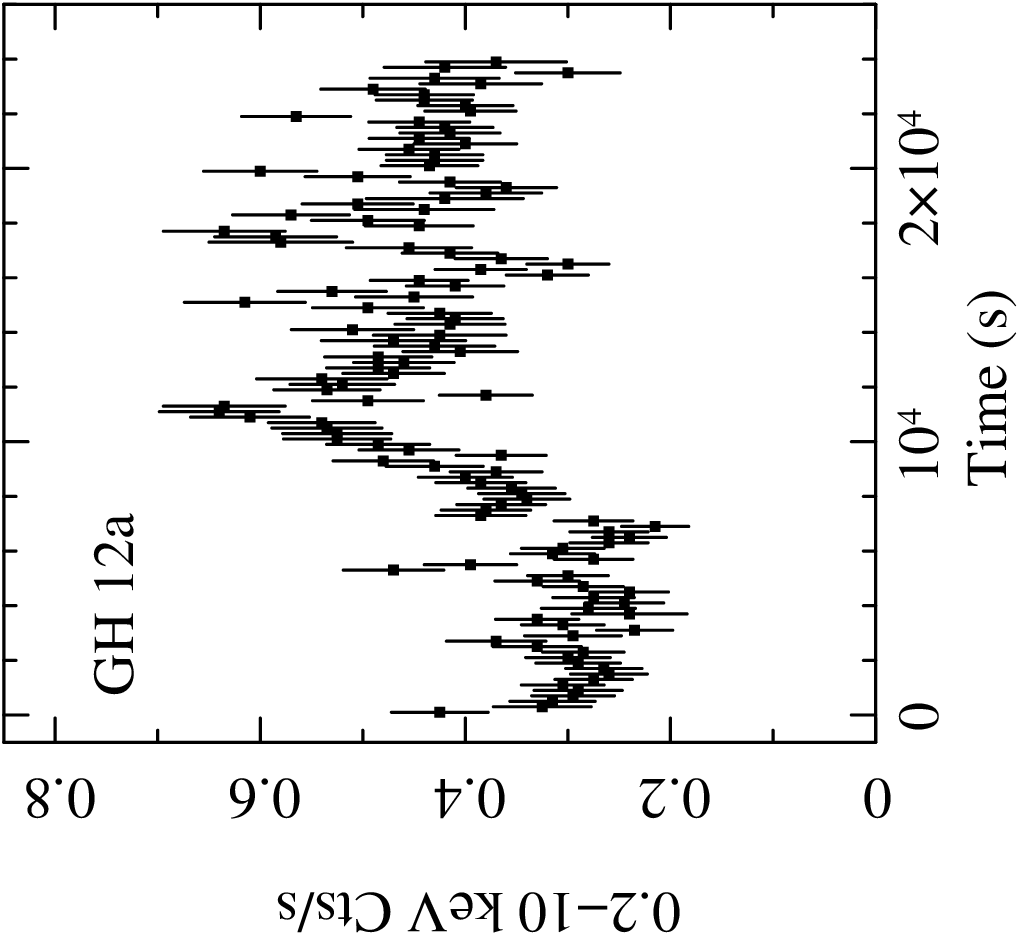}
{\hspace{0.25cm}}
\includegraphics[width=0.35\textwidth,height=0.14\textwidth,angle=-90]{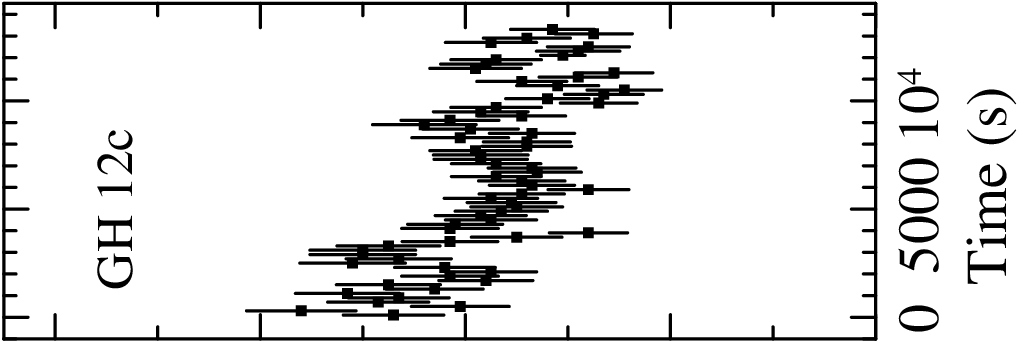}
{\hspace{0.4cm}}
\includegraphics[width=0.35\textwidth,height=0.47\textwidth,angle=-90]{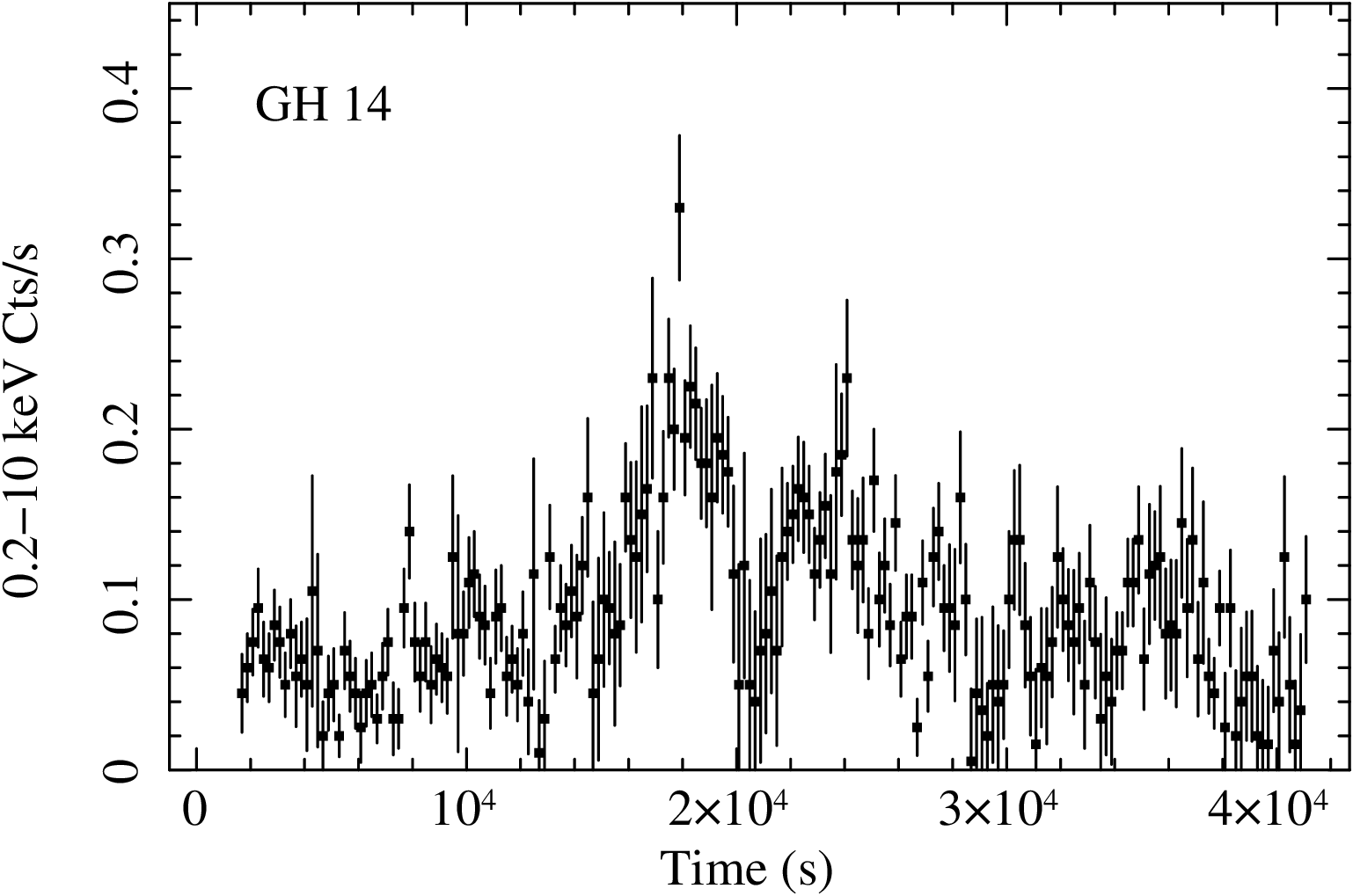}
}}
\caption{Background--subtracted 0.2--10~keV EPIC pn light curves for
  the four objects (common time bin size of 200~s). For GH12, we show
  the light curves from the longest GH12a and GH12c (69~d apart)
  observations only (during the shorter GH~12b observation, the mean
  count rate is $\sim$0.5~Cts/s).}
\label{lcs}
\end{center}
\end{figure*}

\section{X--ray variability}

One of the main goals of our {\it XMM--Newton} observations was to
search for X--ray variability in our mini--sample of IMBHs. Since the
BH mass sets the geometrical scale of the system, and since X--rays
are thought to originate in the innermost few tens gravitational
radii, short timescale variability has to be expected from our IMBH
sample. As a reference, the light crossing time of a sphere of
$50~r_g$ in radius is $\sim$500~s for a $10^6~M_\odot$ BH. Hence,
X--ray variability down to a few hundred seconds (at least) has to be
expected for our sources if, as generally assumed, the X--ray emission in AGNs
originates in the innermost regions of the accretion flow.

In Fig.~\ref{lcs} we show the broadband (0.2--10~keV) background
subtracted light curves for all our objects with a 200~s time--bin, as
obtained from the EPIC pn data. X--ray variability is clearly detected
in all cases down to the bin--size timescale, confirming that the
X--ray emitting region must be compact, of the order of a few tens of
gravitational radii.

One of the most powerful ways to investigate the nature and properties
of X--ray variability (and any time--series) is the analysis of the
power spectral density (PSD), i.e. the distribution of the power of
the signal as a function of frequency. At least one characteristic
frequency (or timescale) can be derived from the PSD of accreting
black hole X--ray light curves.  If the PSD is parameterized as $P(\nu)
\propto \nu^{-\alpha}$, most AGN and X--ray binaries in the high/soft
state have $\alpha \approx 2$ which breaks to $\alpha \approx 1$ below
a characteristic high frequency break $\nu_{\rm H}$. In addition to
the high frequency break, X--ray binaries in the hard and intermediate
states also exhibit a second low frequency break $\nu_{\rm L}$ and,
for frequencies below $\nu_{\rm L}$, the PSD slope becomes
$\alpha\approx 0$. For AGN, convincing evidence for a low frequency
break has been reported in Ark~564 only (M$^{\rm c}$Hardy et al 2007).

\subsection{Excess variance}

In the case of our IMBH observations, the data quality is not high
enough to accurately measure the PSD shape and test for the presence
of reliable high frequency breaks. However, a comparison between the
IMBH X--ray variability and that of more massive AGN can be obtained
by the analysis of the so--called excess variance $\sigma^2_{\rm
  NXS}$, which is the integral of the power spectrum over a given
frequency window $\Delta\nu$, defined by the light curve duration and
time--bin (Nandra et al 1997; Edelson et al 2002; Vaughan et al 2003).

Early results pointed out an anti--correlation between excess variance
and AGN luminosity (Green, M$^{\rm c}$Hardy \& Letho 1993; Lawrence \&
Papadakis 1997; Nandra et al 1997).  Later study by Lu \& Yu (2001)
however outlined an underlying anti--correlation between
$\sigma^2_{\rm NXS}$ and black hole mass on variability timescales of
$\sim$~1 day, a result later extended to longer timescales by
Papadakis (2004). More recently, O'Neill et al (2005) have computed
$\sigma^2_{\rm NXS}$ for a large sample of AGN of known black hole
mass observed with {\it ASCA}, confirming the general
anti--correlation seen before, and pointing out that the observed
behaviour can be explained by assuming a universal form of the PSD
with both a low and a high frequency break and by further assuming
that the PSD amplitude is uniform (i.e. the same in all sources) and
that the frequency breaks (most notably $\nu_{\rm H}$) are
proportional to $M^{-1}_{\rm BH}$ (see e.g. Papadakis 2004; Markowitz
et al 2003; O'Neill et al 2005). While such model reproduces the
$\sigma^2_{\rm NXS}$--$M_{\rm BH}$ relationship well, the scatter in
the relationship is too large to compute errors on the parameters
(O'Neill et al 2005). In fact, it must be stressed that the above
assumptions are only zeroth--order approximations because the PSD
amplitude is known to vary significantly from source to source
(e.g. Uttley \& M$^{\rm c}$Hardy 2005) and $\nu_{\rm H}$ has been
convincingly shown to be not only a
function of the black hole mass, but also of the accretion rate 
(M$^{\rm c}$Hardy et al 2006). Moreover, the low--frequency break
$\nu_{\rm L}$ affects the low--mass end curvature in the relationship
but is most likely outside the observable frequency window for typical
X--ray observations of AGN with the exception of Ark~564; M$^{\rm
c}$Hardy et al. (2007), and one object only is clearly not enough to
infer the general scaling between the low and high frequency breaks in
AGN.

\begin{figure}
\begin{center}
\includegraphics[width=0.33\textwidth,height=0.45\textwidth,angle=-90]{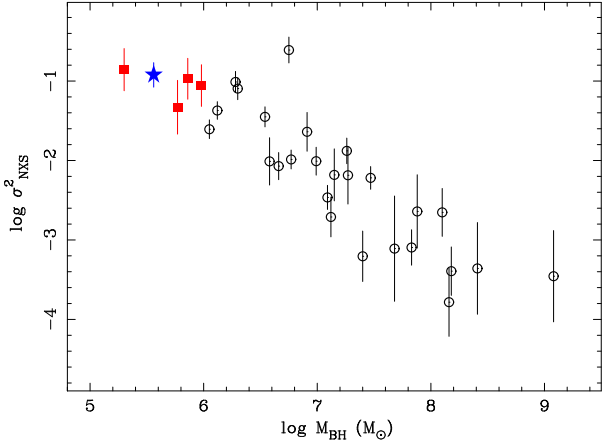}
\caption{ {\bf Top:} Excess variance as a function of black hole
  mass. Filled squares represent our mini--sample of IMBHs, the
  filled star is NGC~4395, and open circles for a sample of
  radio--quiet Seyfert 1
  galaxies observed with {\it XMM--Newton} (see Table~5 and text for details). 
}
\label{sig}
\end{center}
\end{figure}

\begin{table}
\begin{center}
  \caption{Black hole mass and excess variance for a sample of 27
  radio quiet Seyfert~1 galaxies (including also NLS1 galaxies and 1
  IMBH, e.g. NGC~4395) and for our 4 IMBH.}
\begin{tabular}{lccc}          
  \hline
  ID &$\log$M$_{\rm{BH}}$ & $\log \sigma^2_{\rm{NXS}}$ & Ref$^\dagger$\\  
  \hline
  GH~1  & $5.86$ & $-0.97\pm 0.26$ & [1]\\ 
  GH~8  & $5.77$ & $-1.33\pm 0.34$ & [1]\\ 
  GH~12 & $5.98$ & $-1.06\pm 0.26$ & [1]\\ 
  GH~14 & $5.30$ & $-0.86\pm 0.27$ & [1]\\    
\hline
NGC~4395  & $5.56$ & $-0.92 \pm 0.15$ & [2]\\  
Mrk~766    & $6.05$ & $-1.61\pm  0.12$ & [3]\\
Ark~564   & $6.12$ & $-1.37\pm  0.11$ & [4] \\  
NGC~4051    & $6.28$ & $-1.01\pm  0.13$ & [5]\\  
1H~0707--495 & $6.30$ & $-1.10\pm  0.14$ & [3]\\
MCG--6-30-15  & $6.54$ & $-1.45\pm 0.13$ & [6]\\  
MS~2254--36  & $6.58$ & $-2.01\pm 0.30$ & [7]\\ 
WAS~61& $6.66$ & $-2.07\pm  0.17$ &[7]\\
IRAS~13224--3809 & $6.75$ & $-0.61\pm  0.16$ & [3] \\  
IC~4329A    & $6.77$ & $ -1.99\pm 0.12$ & [8]\\  
TonS~180    & $6.91$ & $-1.64\pm 0.24$ & [3]\\  
NGC~4593    & $6.99$ & $-2.01\pm 0.18$ & [9]\\  
NGC~7469    & $7.09$ & $-2.46\pm 0.15$ & [5]\\  
NGC~4151    & $7.12$ & $-2.71\pm  0.25$ & [5]\\  
Mrk~335    & $7.15$ & $-2.18\pm  0.33$ & [5]\\  
IZw1    & $7.26$ & $-1.88\pm 0.16$ & [3]\\  
Mrk~478    & $7.27$ & $-2.19\pm  0.36$ & [3]\\  
Mrk~110    & $7.40$ & $-3.21\pm 0.32$ & [5]\\  
NGC~3783 & $7.47$ & $-2.21\pm  0.14$ & [5]\\  
Mrk~590    & $7.68$ & $-3.11\pm 0.66$ & [5]\\  
NGC~5548    & $7.83$ & $-3.09\pm 0.22$ & [5]\\  
PG~1211+143    & $7.88$ & $-2.64\pm 0.46$ & [3]\\  
Mrk~841    & $8.10$ & $-2.65\pm 0.30$ & [8]\\  
Mrk~509    & $8.16$ & $-3.78\pm 0.43$ & [5]\\  
Ark~120    & $8.18$ & $-3.39\pm  0.31$ & [5]\\  
Fairall~9    & $8.41$ & $-3.36\pm 0.58$ & [5]\\  
HE~1029--1401    & $9.08$ & $-3.46\pm 0.57$ & [3]\\  
\hline 
\end{tabular}
\parbox{3.28in} { $^\dagger$ References for
black hole masses are: 
 [1] Greene \& Ho (2004); [2] Peterson et al (2005); [3] Wang et
 al. (2004); [4] Botte et al. (2004); [5] Peterson et al. (2004); [6]
 M$^{\rm c}$Hardy et al. (2005); [7] Grupe et al. (2004); [8] Woo \&
 Urry (2002); [9] Denney et al. (2006)
}
\end{center}
\end{table}

Because of the above limitations, it is not our purpose here to repeat
the O'Neill et al analysis (as we cannot constrain the parameters of
the model any better), but in order to compare the IMBH variability
properties with more massive objects, we consider a sample of 27
radio--quiet Seyfert 1 galaxies observed with {\it XMM--Newton} plus
our 4 IMBHs and compute the excess variance by using light--curve
segments of equal duration ($T=20$~ks) and equal time bin size
($\delta T = 500$~s) in the 2--10~keV band for all sources. The sample
was selected from the available sources (with known/estimated black
hole mass) in the {\it XMM--Newton} public archive with observations
longer than 20~ks. The sample is by no means homogeneous nor complete
and results are reported here for comparison with the IMBH only. If
observations longer than 20~ks (or multiple observations) are
available for a given source, the excess variance has been estimated
as the mean over the excess variances obtained from the 20~ks
intervals and errors are computed taking into account the measurement
errors, and also the uncertainties associated with red--noise (Vaughan
et al. 2003). In the case of GH~12, we used observation GH~12a only,
and we imposed a slightly less conservative background cut than for
the spectral analysis in order to obtain a 20~ks segment of
the 2--10~keV light curve (we checked that the different cut does not
affect the excess variance significantly).

The excess variances and black hole masses used here
are reported in Table~5. In Fig.~8, we show the resulting $\log
\sigma^2_{\rm NXS}$--$\log M_{\rm BH}$ relationship, showing that the
excess variance of our IMBHs (filled squares) smoothly joins with that
of more massive sources. Such behaviour is consistent with that
generally predicted by the universal PSD model, although with the
model limitations discussed above (see e.g. Fig.~2 in O'Neill et al
2005) and we conclude that the IMBH X--ray variability properties do
not differ significantly from those of more massive Seyfert
galaxies. Only extending the available data to black hole masses below
$\sim 10^5~M_\odot$ would better define the relationship in the most
crucial low--mass end where, according to the universal PSD model, the
excess variance is expected to significantly deviate from the
anti--correlation (and actually to nearly saturate if, as results from
Ark~564 indicate, the low frequency break in AGN is $\nu_{\rm L} \leq
10^{-3}\times \nu_{\rm H}$).

\section{Discussion}

We have observed with {\it XMM--Newton} 4 IMBH from the original
sample of 19 objects selected by Greene \& Ho (2004) from the
SDSS. The X--ray properties of the IMBH considered here do not
significantly differ from those of their more massive counterparts
such as the PG quasars. The IMBHs could represent a population of not
yet fully grown super--massive black holes and, if in a rapid growing
phase, their accretion properties would be expected to significantly
depart from those of standard and fully--grown accreting supermassive black
holes. However, as already suggested by their Eddington ratios (high
because of the sample selection strategy, but not that unusual), their
accretion properties are in all respect similar to those of more
massive AGN and quasars.

The X--ray spectra of our objects are characterized by the same X--ray
slope as PG quasars in both the hard and soft X--ray bands. No
(neutral) absorption in excess of the Galactic one is observed in any
of the objects. Three out of four objects exhibit a soft excess with
the same properties (uniform temperature and strength) as larger black
hole mass samples (standard Seyfert~1 and quasars). This is surprising
if one considers that the small black hole mass of the IMBHs should
allow thermal emission from the accretion disc to dominate the soft
X--ray band, producing soft X--ray spectra that should significantly
depart from those of more massive objects in which the disc
quasi--blackbody emission is expected to contribute in the UV
only. This is however not observed, and if the soft excess is
interpreted as thermal disc emission, the inferred temperature in AGN
with black hole masses of $\leq 10^6~M_\odot$ is the same as that of
quasars with black hole masses two (or more) orders of magnitude
larger. So far, the so--called soft excess problem was mainly defined
as the fact that we observe soft X--ray emission in accreting
supermassive black hole which is too hot (and too uniform) to be
reasonably associated with thermal disc emission. Here, we point out
that the problem is even more puzzling: when we observe sources in
which thermal disc emission should largely extend into the soft X--ray
band (such as the IMBH), we see exactly the same spectral shape as in
more massive AGN and, if these IMBH were rotating (Kerr), the inferred
temperature would even be too cold to be associated with the accretion
disc. 

\subsection{Smeared absorption}

Gierli\'nski \& Done (2004) proposed that the soft excess is an
artifact due to absorption (of an intrinsic steep power law spectrum)
by partially ionized gas in the form of a highly relativistic wind.
Since the observed smoothness of the soft excess is not consistent
with the expected sharp absorption features, the model requires the
gas to be characterized by strong velocity gradients to smear them out
via Doppler smearing. The large velocity gradients, most likely imply
that the wind is launched very close to the black hole. 

As mentioned by Schurch \& Done (2006), the wind model described above
is unlikely to be a standard UV--line driven disc wind, because such
winds are preferentially produced at $i>75^\circ$ of the axis, which
is in contrast with the observational fact that most low--inclination
Seyfert 1 galaxies exhibit a soft excess. Moreover, the mass outflow
rate ($\dot M_{\rm out}$) inferred from the large final velocities
required by the model is
generally hundreds times larger than the mass accretion rate ($\dot
M_{\rm acc}$) in among the most luminous X--ray sources in the
Universe (but see the idea of a failed wind proposed by Schurch \&
Done 2006). More recently, Schurch \& Done (2007) have shown that when
a more detailed smearing profile is adopted, the shape of the model
does not describe well the observed smoothness of the soft excess,
casting severe doubts on the overall model.

In addition, here we have shown that, even if the smeared absorption
model can reproduce the spectral shape, the uniform soft excess
strength is obtained only if a tight correlation between column
density and ionization parameter is enforced. The $N_H-\xi$ correlation
inferred from fitting the smeared absorption model to X--ray data is
however not a natural product of photo--ionization and is most likely
spurious, either due to a degeneracy in the parameter space (which
implies the model does not provide any constraint on the absorber
properties) or, as mentioned above, to the need of producing a uniform
soft excess strength in sources which do require different column densities.

\subsection{Disc reflection}

The smearing of the reflection features by high velocities is a
natural consequence of the presence of an inner accretion disc.
Undoubtedly, an inner accretion disc must exists in sources accreting
above a few per cent of the Eddington ratio and thus, unlikely the
smeared absorption model, the disc reflection model does not introduce
any new (so far unobserved, if not for the soft excess interpretation
itself) component to explain the soft excess. The reflection
interpretation just makes use of the basic ingredients of accretion
theory, a powerful primary source of hard X--rays, and an optically
thick accretion disc which reprocesses the irradiating flux into a
reflection spectrum. An additional external absorber may obviously
mask this contribution and this may well be the case in some most
extreme sources. However, since disc reflection only requires the
presence of an inner accretion disc and of an illuminating X--ray
source, it is an unavoidable component in any reasonable accretion
model for radiatively efficient AGN (and X--ray binaries). In fact
(see e.g.  Middleton, Done \& Gierli\'nski 2007) when the smeared
absorption model is applied to PG quasars and NLS1s, it always does
require also a disc reflection component. If the disc reflection
component alone can account for the spectral shape and variability of
the sources, it is difficult to understand the need for an additional
smeared absorber with problematic launching mechanism, extreme mass
outflow rate, and rather unnatural relationship between column density
and ionization. 

Since disc reflection invokes the presence of the
accretion disc down to the last stable orbit, thermal emission from
the disc should also be present and, given the small black hole masses
and relatively high accretion rates in our IMBHs, such emission should
peak in the soft X--rays. Indeed, when a multicolor disc blackbody is
included in the spectral models, a solution in which the soft excess
is partly due to reflection and partly to disc blackbody with the
theoretically expected temperature is found. However, we point out
that if thermal disc emission is present in the soft X--ray spectrum
of these objects, its luminosity is far below the expected one for the
given temperatures not only when reflection is invoked, but even when
all the soft excess is modelled as a pure disc blackbody.

It is actually difficult to distinguish between the two competing
models (smeared absorption and disc reflection) by using spectral and
variability information in the limited band--pass of {\it XMM--Newton}
(0.3--10~keV band) even in bright, well observed sources. Moreover,
our mini--sample of small mass BH is not well suited to investigate
the problem due to the relatively low quality of the data.
High--energy data are crucial to test whether the X--ray spectrum
above 10~keV is just the high--energy unabsorbed tail of the intrinsic
power law continuum, or whether it is instead characterized by the
presence of a Compton hump around 20--30~keV, related to X--ray
reflection. While signatures for the presence of X--ray reflection
from the inner accretion disc are sometimes detected, the number of
well observed sources above 10~keV is still too small to draw any
clear--cut conclusion (e.g. Miniutti et al 2007; Reeves et al
2007). Observations with the {\it Suzaku} X--ray mission and with
future missions such as Simbol--X (Ferrando et al. 2006) will most
likely play a crucial role in that direction in the near future.

\subsection{X--ray variability}

Given their relatively small black hole mass (i.e. small size), the
observed IMBH are not surprisingly among the most variable in
X--rays. In particular, their excess variance $\sigma^2_{\rm NXS}$ is
among the largest obtained from AGN X--ray light curves. Our
observations begin to fill a relatively poorly explored range of black
hole masses in the $\sigma^2_{\rm NXS}$--$M_{\rm BH}$ relationship and
show that the X--ray variability properties of IMBH smoothly join with
those of more massive Seyfert galaxies. The $\sigma^2_{\rm
NXS}$--$M_{\rm BH}$ is consistent with a simple (well known)
anti--correlation that can be explained with a universal PSD model in
which the break frequencies (most importantly the high--frequency one)
scale with $M^{-1}_{\rm BH}$. However, such functional form for
$\nu_{\rm H}$ is known to be only a zeroth--order approximation (see
M$^{\rm c}$Hardy et al 2006) and the universal PSD model suffers also
for other uncertainties that do not allow to properly account for the
scatter in the relationship. Future longer observations of these and
other IMBH in X--rays would allow to explicitly search for the high
frequency break in the PSD, enabling us to extend the $\nu_{\rm
H}$--$M_{\rm BH}$--$\dot M_{\rm{acc}}$ relationship pointed out by
M$^{\rm c}$Hardy et al (2006) to the most crucial black hole mass
range of $M_{\rm BH} \leq 10^6~M_\odot$.

\section*{Acknowledgments}

Based on observations obtained with XMM-Newton, an ESA science mission
with instruments and contributions directly funded by ESA Member
States and NASA. We would like to thank the referee for her/his
suggestions that improved our work. GM thanks the UK STFC and the
French CNRS for partial support. GM also thanks the Spanish Ministerio
de Ciencia e Innovaci\'on and CSIC for support through a Ram\'on y
Cajal contract. ACF thanks the Royal Society for support.

\end{document}